%% file: main_pvis.tex
\documentclass[preprint]{vgtc}               % preprint
\ifpdf%                                % if we use pdflatex
  \pdfoutput=1\relax                   % create PDFs from pdfLaTeX
  \usepackage{graphicx}                % allow us to embed graphics files
  \DeclareGraphicsExtensions{.pdf,.png,.jpg,.jpeg} % for pdflatex we expect .pdf, .png, or .jpg files
\else%                                 % else we use pure latex
  \usepackage{graphicx}                % allow us to embed graphics files
  \DeclareGraphicsExtensions{.eps}     % for pure latex we expect eps files
\fi%

\graphicspath{{figures/}{pictures/}{images/}{./}} % where to search for the images

\usepackage{microtype}                 % use micro-typography (slightly more compact, better to read)
\PassOptionsToPackage{warn}{textcomp}  % to address font issues with \textrightarrow
\usepackage{textcomp}                  % use better special symbols
\usepackage{mathptmx}                  % use matching math font
\usepackage{times}                     % we use Times as the main font
\usepackage{cite}                      % needed to automatically sort the references
\usepackage{tabu}                      % only used for the table example
\usepackage{booktabs}                  % only used for the table example
\usepackage{amsmath}
\usepackage{gensymb}
\usepackage{amssymb}
\usepackage{enumitem}
\usepackage[usenames, dvipsnames]{color}
\usepackage[dvipsnames]{xcolor}
\usepackage[caption=false]{subfig}
\usepackage{calrsfs}
\usepackage[keeplastbox]{flushend}

\onlineid{4722}

\vgtccategory{Evaluation}
\vgtcinsertpkg

\preprinttext{\scriptsize To appear in the proceedings of the IEEE Pacific Visualization Symposium 2020}

\title{A Study of Mental Maps in Immersive Network Visualization}

\author{%
Joseph Kotlarek\thanks{e-mail: jtkotlarek@ucdavis.edu}\\ \scriptsize University of California, Davis%
\and Oh-Hyun~Kwon\thanks{e-mail: kw@ucdavis.edu}\\ \scriptsize University of California, Davis%
\and Kwan-Liu~Ma\thanks{e-mail: ma@cs.ucdavis.edu}\\ \scriptsize University of California, Davis%
\and Peter~Eades\thanks{e-mail: peter.eades@sydney.edu.au}\\ \scriptsize University of Sydney%
\vspace{1em}
\and Andreas~Kerren\thanks{e-mail: andreas.kerren@lnu.se}\\ \scriptsize Linnaeus University%
\and Karsten~Klein\thanks{e-mail: karsten.klein@uni-konstanz.de}\\ \scriptsize University Konstanz%
\and Falk~Schreiber\thanks{e-mail: falk.schreiber@uni-konstanz.de}\\ \scriptsize University of Konstanz%
}

\abstract{\input{tex/abstract.tex}} % end of abstract

\CCScatlist{
  \CCScatTwelve{Human-centered computing}{Visu\-al\-iza\-tion}{Visu\-al\-iza\-tion techniques}{Graph drawings};
  \CCScatTwelve{Human-centered computing}{Visu\-al\-iza\-tion}{Empirical studies in visualization}{}
}

\newif\ifthesis
\thesisfalse

\begin{document}


\input{tex/intro.tex}
\input{tex/related_work.tex}
\input{tex/experiment.tex}
\input{tex/result.tex}
\input{tex/discussion.tex}
\input{tex/conclusion.tex}

\acknowledgments{This research is sponsored in part by the U.S. National Science Foundation
through grants IIS-1741536 and IIS-1528203.}

\bibliographystyle{abbrv}

\bibliography{tex/references}
\end{document}

%% file: tex/abstract.tex
The visualization of a network influences the quality of the mental map that the viewer develops to understand the network. 
In this study, we investigate the effects of a 3D immersive visualization environment compared to a traditional 2D desktop environment on the comprehension of a network’s structure. 
We compare the two visualization environments using three tasks---interpreting network structure, memorizing a set of nodes, and identifying the structural changes---commonly used for evaluating the quality of a mental map in network visualization. 
The results show that participants were able to interpret network structure more accurately when viewing the network in an immersive environment, particularly for larger networks. 
However, we found that 2D visualizations performed better than immersive visualization for tasks that required spatial memory.

%% file: tex/intro.tex
\firstsection{Introduction}
\maketitle
\label{sec:intro}
Networks are commonly used for representing relational data found in diverse domains, such as social networks, biological networks, communication networks, and power grids.
A non-graphical representation of a network makes it very difficult for humans to understand its structure.
Thus, network visualization is widely used to make sense of the structural information in the data.
The visualization of a network helps to form a \emph{mental map} of the data, which is the internal representation of the information inside the viewer's mind that is formed by looking at the visualization.

Most existing studies on mental maps in the context of network visualization have focused on dynamic network data~\cite{Archambault18Shonan, Archambault13, Archambault12, Archambault11, Purchase07, North96, Misue95, Purchase08}.
More specifically, a significant body of the above research has investigated the effect of different layout methods on \emph{mental map preservation},
which is often interpreted as \emph{drawing stability}~\cite{Archambault18Shonan}: minimizing positional changes of nodes and edges of the network in the visualization between successive time periods.
While mental maps in network visualization have frequently been studied in the context of dynamic network data, the quality of the mental map is also important for interactive visualization~\cite{Archambault18Shonan}, as the interactions performed by the user often changes the visual representation of the data.
In the context of this work, we define the quality of a mental map as that which helps the viewer form an accurate understanding of the structure, connectivity, and content of the network.
The quality of the mental map is thus influenced by the visual representation of the network.

When designing interactive visualizations, one needs to consider various factors such as input devices and display environments.
In recent years, new input and display technologies (e.g., large wall displays, table top displays with touch and pen input devices, and immersive 3D virtual/mixed reality environments) have become ubiquitous enough that researchers can consider these technologies for general visualization designs~\cite{Dwyer18}.
Immersive visualization and analytics is an emerging field that aims to leverage the many benefits of mixed reality technology to information visualization, such as \emph{spatial immersion}, \emph{multi-sensory presentation}, and \emph{increased engagement}~\cite{Dwyer18}. 
In the context of network visualization,
immersive technologies have been used for facilitating novel visualization and interaction techniques~\cite{Kwon15, Kwon16, Greffard12, Marriott18, Huang17}.

In this study, we investigate the effects of an immersive 3D visualization environment on the quality of a mental map for comprehending network data.
Specifically, we conduct a controlled user study to compare (1) a 3D immersive visualization environment with a head-mounted display and hand-tracker controllers and (2) a traditional 2D desktop environment with standard monitor and mouse.
We study three tasks that are commonly used for evaluating the quality of a mental map in network visualization \cite{Archambault13}: interpreting network structure, memorizing a set of nodes, and identifying the structural changes.

The results show that participants completed the tasks faster, and recalled changes to the network more accurately in the desktop environment.
However, they interpreted network structures more accurately in the immersive environment, and reported the task as easier to complete in the immersive environment.
An overwhelming majority of participants (85\%) favored the immersive environment over the desktop environment.
The results indicate that immersive environments mitigate confusion caused by edge crossings and node overlaps that are commonly found in 2D representations, while traditional 2D representations provide a better spatial overviews.
We discuss the implications of these results for future network visualization designs in immersive environments.

%% file: tex/related_work.tex
\section{Related Work}
Immersive environments, such as CAVEs \cite{cave2} and head-mounted displays (HMDs), have been proven effective for various applications. 
Ball and North~\cite{Ball05} have shown that high resolution tiled displays improve perception and navigation for visual tasks.
Mania and Chalmers~\cite{Mania01} have studied memory in immersive and non-immersive spaces and found that immersion significantly improved recall for simple memory tasks. 
Krokos et al.~\cite{Krokos18} have recently shown that recall can be improved with virtual reality (VR) through techniques such as the memory palace metaphor.
Additionally, Kwon et al.~\cite{Kwon16} have shown that immersive network visualization has clear potential user performance on network interpretation tasks.
While immersive technology may have been eschewed in the past, studies such as these in tandem with improvements in stereoscopic displays mark the clear entrance of immersive visualization into the realm of practicality~\cite{Huang17}.

While keyboard and mouse is the ubiquitous standard for desktop interaction, recent studies show intuitive interaction in VR environments can outperform traditional interaction systems, such as the work by Huang et al.~\cite{Huang17} to create a VR gesture system for network visualization. This is no surprise, as B\"uschel et al.~\cite{Buschel18} state, intuitive and low-effort interaction is key to leveraging the benefits of immersion, and keeping users invested in the environment. With different methods of interaction, however, different methods of visualization are also in order.

There has been a significant amount of research adapting virtual reality and other immersive technology to applications in a wide breadth of fields, including biomedical imaging~\cite{Usher18}, scientific visualization~\cite{Van00,Bryson96,DeRidder15}, education~\cite{Zyda05,Hall98,Radu17,Wickens92}, and collaboration~\cite{Cordeil17}.
These studies primarily focused on the application in their respective fields. 
While immersive scientific visualization was quick to establish itself, the impact of immersion on abstract data visualization remains largely unexplored. 
Indeed, very few studies have considered the mental map in immersive network visualization. 
In the past year, however, more research into immersive visualizations with abstract data has been completed.
Drogemuller et al.~\cite{Drogemuller18} evaluate navigation techniques for 3D network visualizations in virtual reality.
Greffard~et~al.~\cite{Greffard12} introduced an immersive visualization designed to preserve the mental map. 
The work in this paper differs from the works above in that we instead investigate the impact of immersion on the mental map. 

Mental maps are typically used to measure the quality of a dynamic network layout~\cite{Archambault11, Purchase07, Diehl01, North96, Misue95}, and the importance of mental map preservation in dynamic layouts has been investigated by several studies~\cite{Purchase07, Purchase08}. 
Previous work by Archambault and Purchase~\cite{Archambault12} investigates mental map preservation in a traditional (non-immersive) environment to show it can help users orientation with tasks such as location and path finding. 
Herman et al.~\cite{Herman00} emphasize the importance of considering \emph{predictability} which is also referred to as \emph{preserving the mental map} in dynamic network layouts. 
As mentioned above, immersive environments also improve navigation and orientation~\cite{Ball05,Cordeil17}, but no study has yet combined these techniques.

\begin{figure}[t]
\includegraphics[width=\columnwidth]{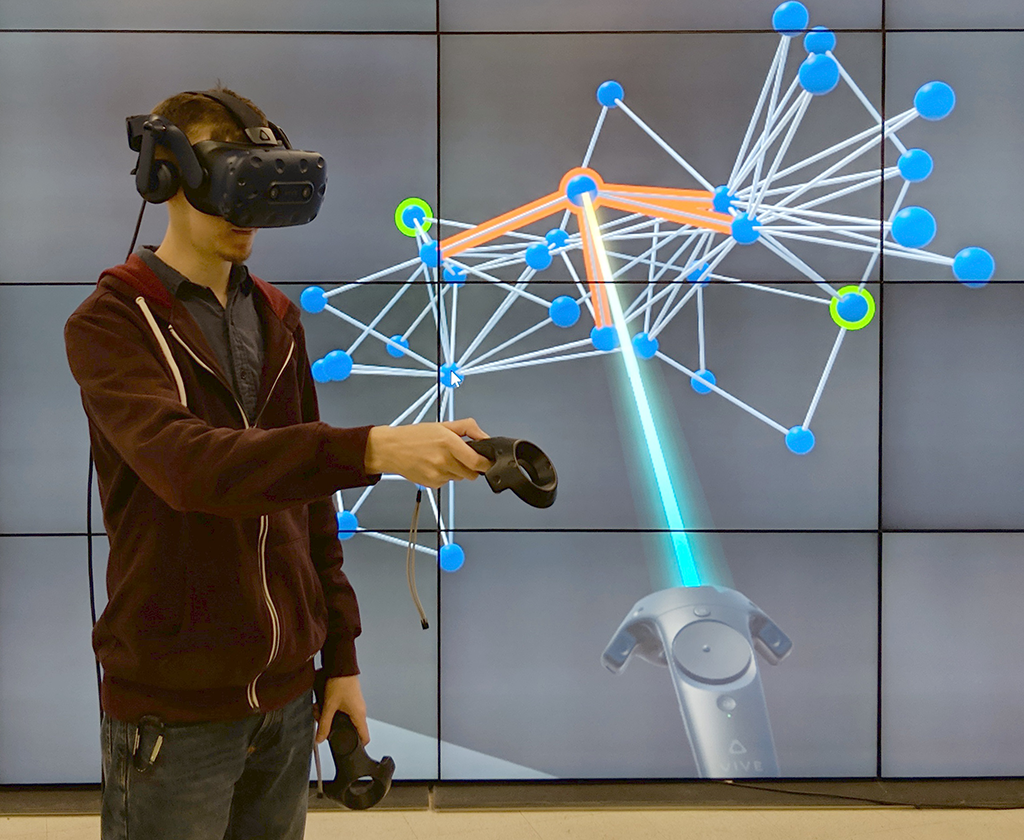}
\caption{A user interacts with an immersive 3D network visualization using a head-mounted display and controllers. 
A virtual laser pointer is implemented for highlighting and selecting nodes in a network.
In addition, users can scale and relocate the network visualization using pinch-to-zoom interaction with both controllers.
The wall display behind the user mirrors the user's view. 
We compare this immersive network visualization with traditional desktop 2D network visualization to determine how it affects users' mental maps.
This figure shows the \emph{karate} network in a 3D layout, which was used in the training session.}
\label{fig:teaser}
\end{figure}

%% file: tex/experiment.tex
\begin{figure}[t]
\includegraphics[width=\columnwidth]{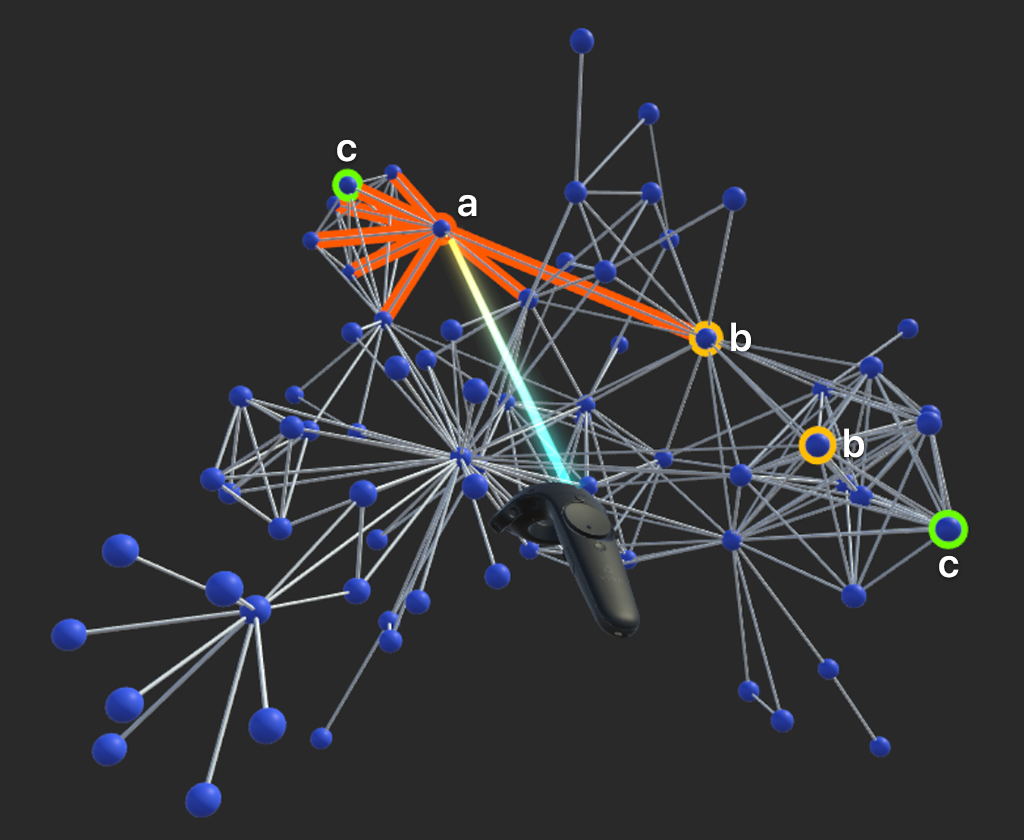}
\caption{Highlighting and selecting nodes.
(a) The orange node is being highlighted over by the user, ready to be selected. 
The edges connected to the highlighted node are also highlighted in orange while the user is pointing it with the controller.
(b) The yellow nodes are the selected nodes by the user. 
(c) The light green nodes are pre-highlighted by the system to show task information to the user.
This figure shows a \emph{path} task.
The goal is to find a shortest path between the light green nodes (c), 
where the nodes in the path needs to be selected (b).
Participant can inspect the connected edges of a node by highlighting the node (a) with the virtual laser pointer.
This figure shows the \emph{lesmis} network in a 3D layout.}
\label{fig:highlight}
\end{figure}

\section{Experiment}
The main purpose of our study is investigating the difference in mental map qualities between two network visualization conditions: immersive 3D visualization and traditional 2D visualization.
We designed a within-participant experiment: 2 \emph{visualization conditions} $\times$ 3 \emph{tasks} $\times$ 3 \emph{networks}. 
The two visualization conditions are compared in terms of \emph{task completion time} and \emph{accuracy}.
This section describes the considerations and design of the experiment.

\subsection{Visualization Conditions}
\label{sec:vis-cond}
There are many different factors in designing a network visualization, such as the layout of the network, color scheme, and interaction techniques.
To remove possible confounding factors, we need to focus on certain key factors that we wish to compare, while keeping other factors the same.
In this study, we focus on the dimensionality (2D and 3D) of a visualization environment. 
For this, we designed the following two visualization conditions:
\begin{itemize}[topsep=3pt, partopsep=0pt, itemsep=3pt, parsep=0pt]
\item \textbf{\emph{2D}}: This condition visualizes a network using a 2D layout in a traditional desktop display. 
In this condition, users use a mouse for interactions: highlight and select nodes, and navigate (pan and zoom) a network. 
\item \textbf{\emph{3D}}: This condition renders a network using a 3D layout in an immersive head-mounted display (HTC Vive Pro). 
Users use controllers for interactions, as shown \autoref{fig:teaser}. 
Room-scale tracking is used for tracking position and rotation of the HMD and controllers.
\end{itemize}

\subsubsection{Layout} 
\begin{figure*}[t]
\captionsetup{farskip=3pt}
\subfloat[\emph{karate} (2D)]{\includegraphics[width=0.245\textwidth]{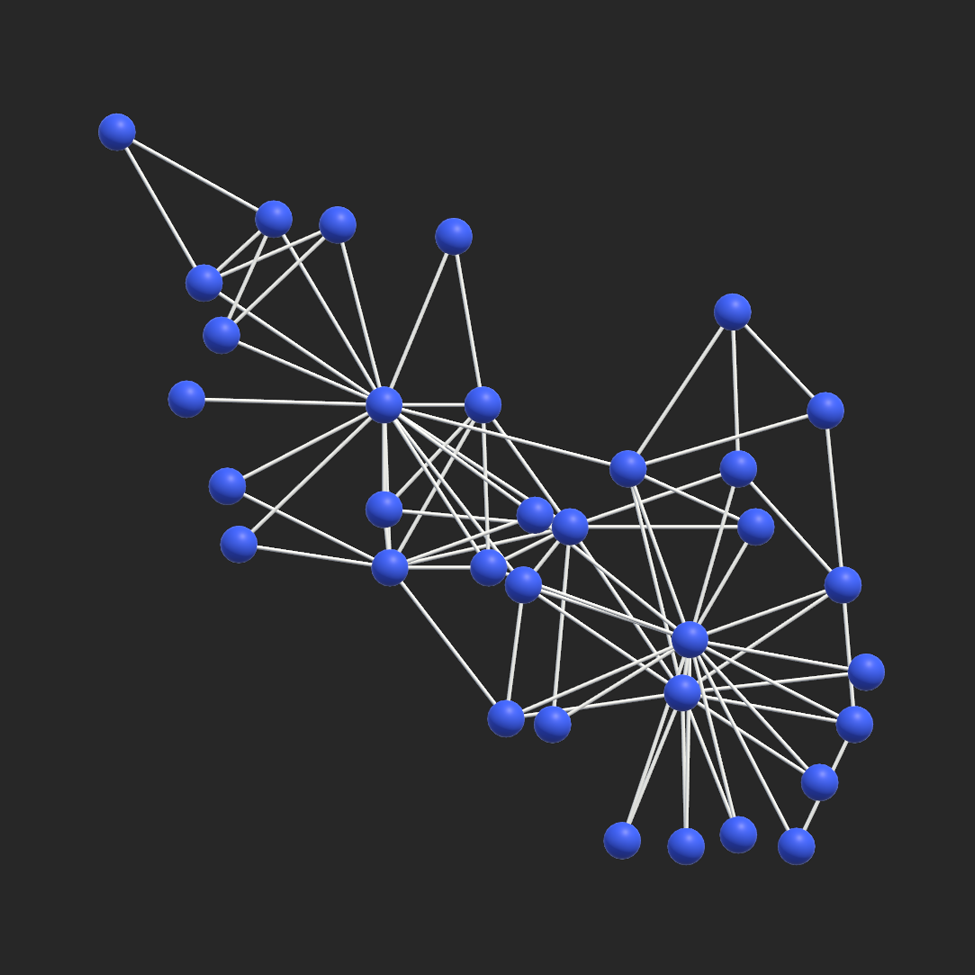}}\hfill
\subfloat[\emph{karate} (3D)]{\includegraphics[width=0.245\textwidth]{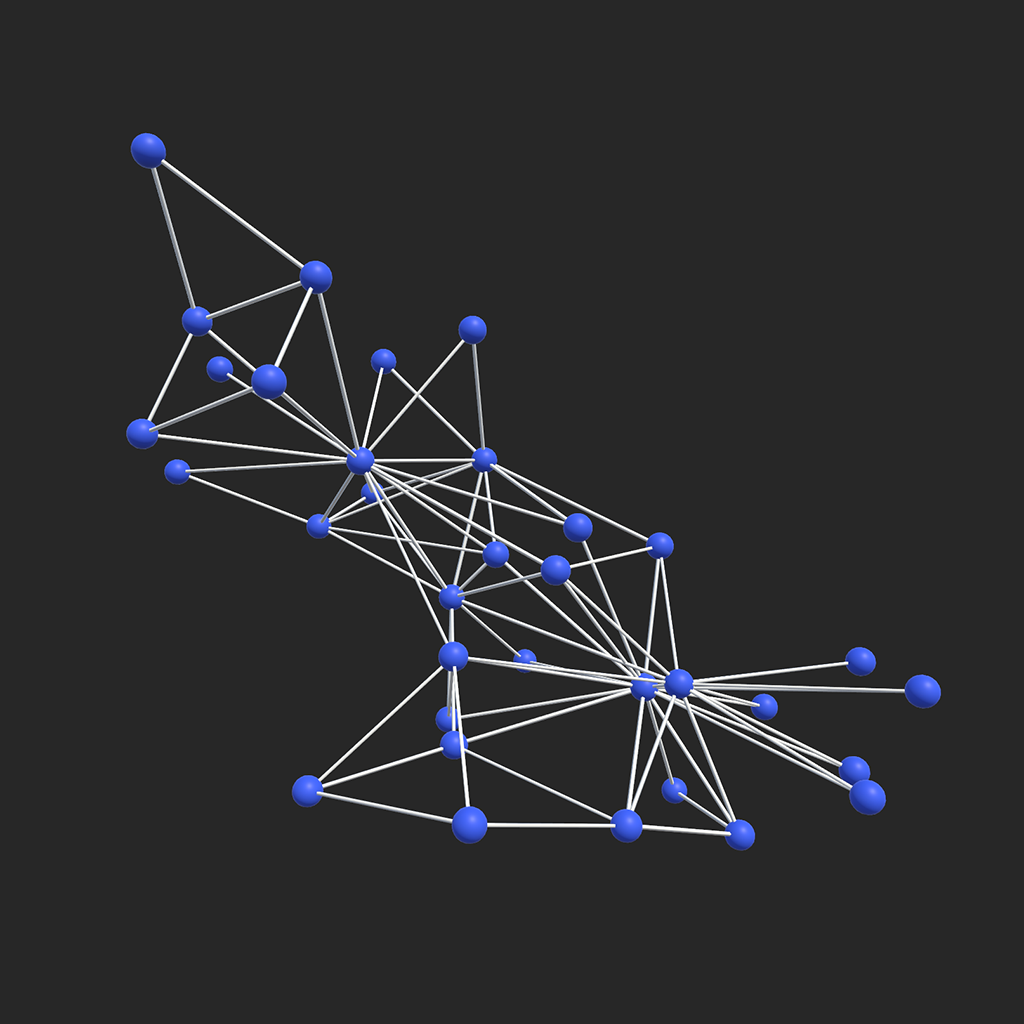}}\hfill\hfill\hfill
\subfloat[\emph{lesmis} (2D)]{\includegraphics[width=0.245\textwidth]{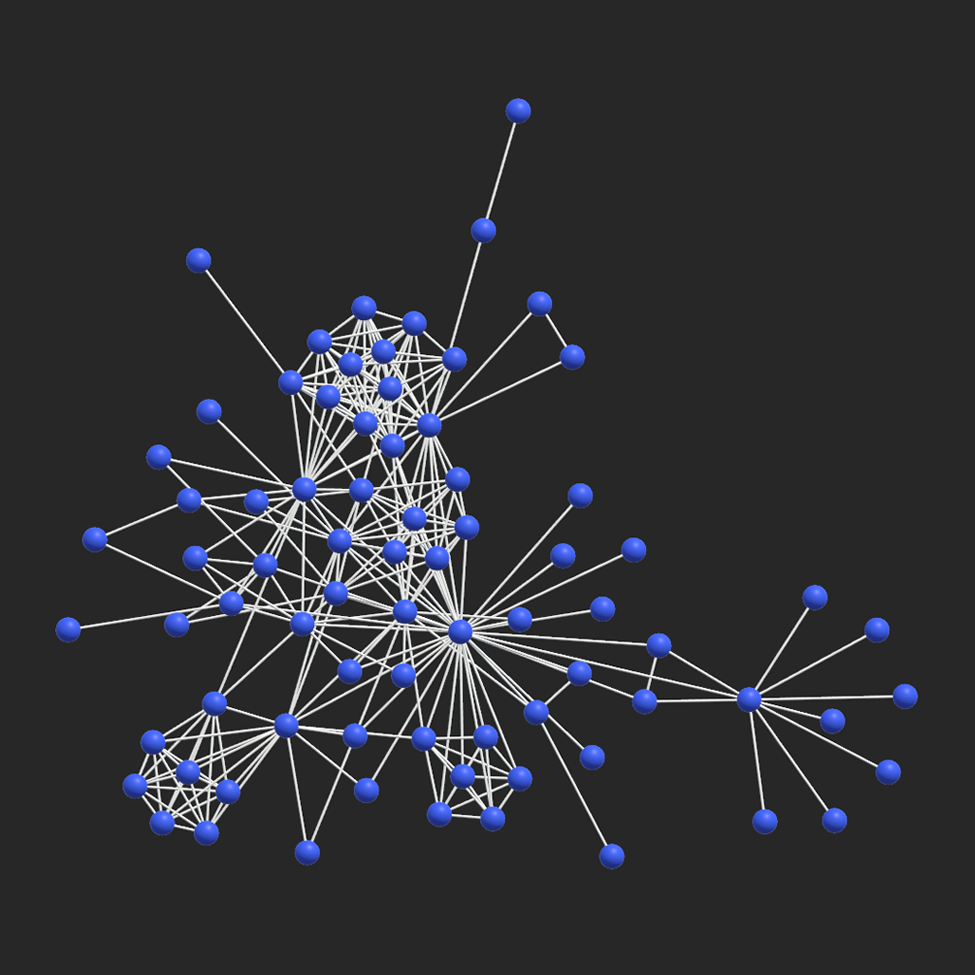}}\hfill
\subfloat[\emph{lesmis} (3D)]{\includegraphics[width=0.245\textwidth]{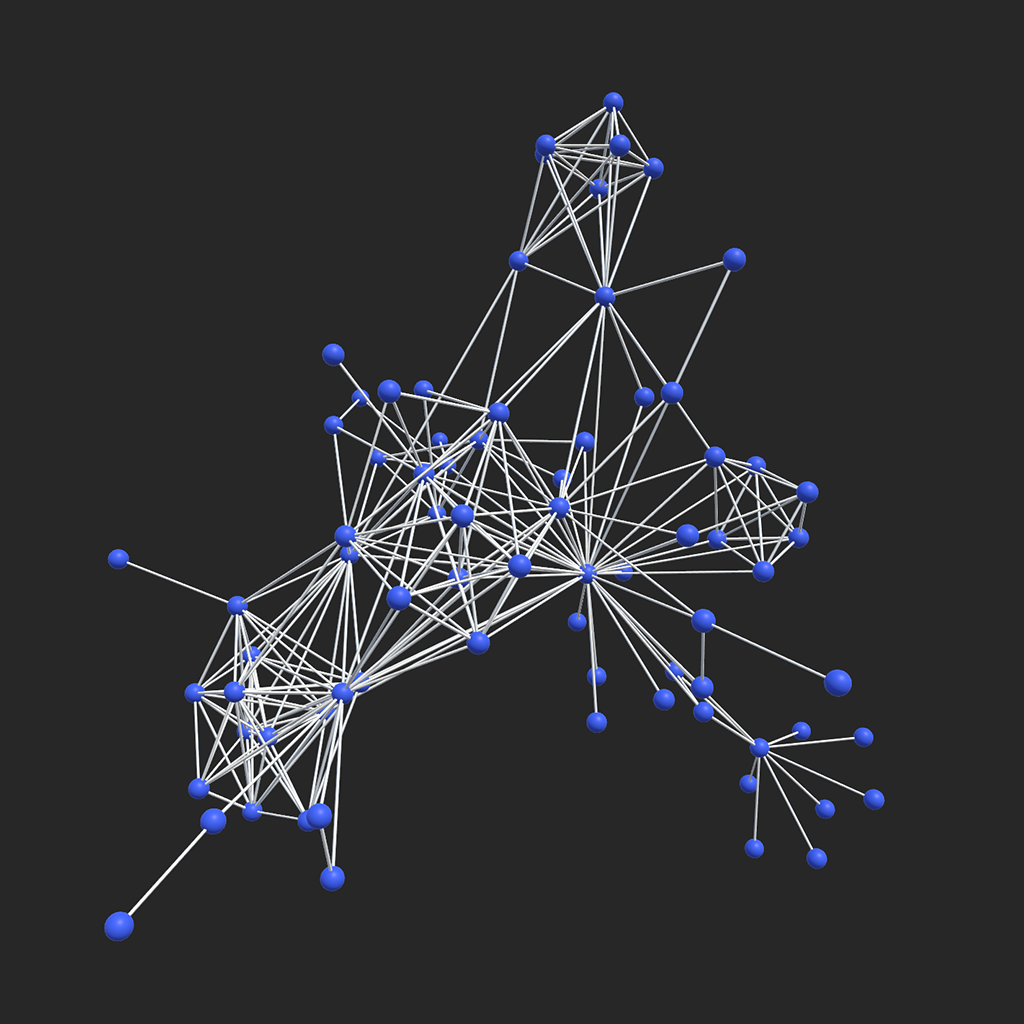}}\\
\subfloat[\emph{netsci} (2D)]{\includegraphics[width=0.245\textwidth]{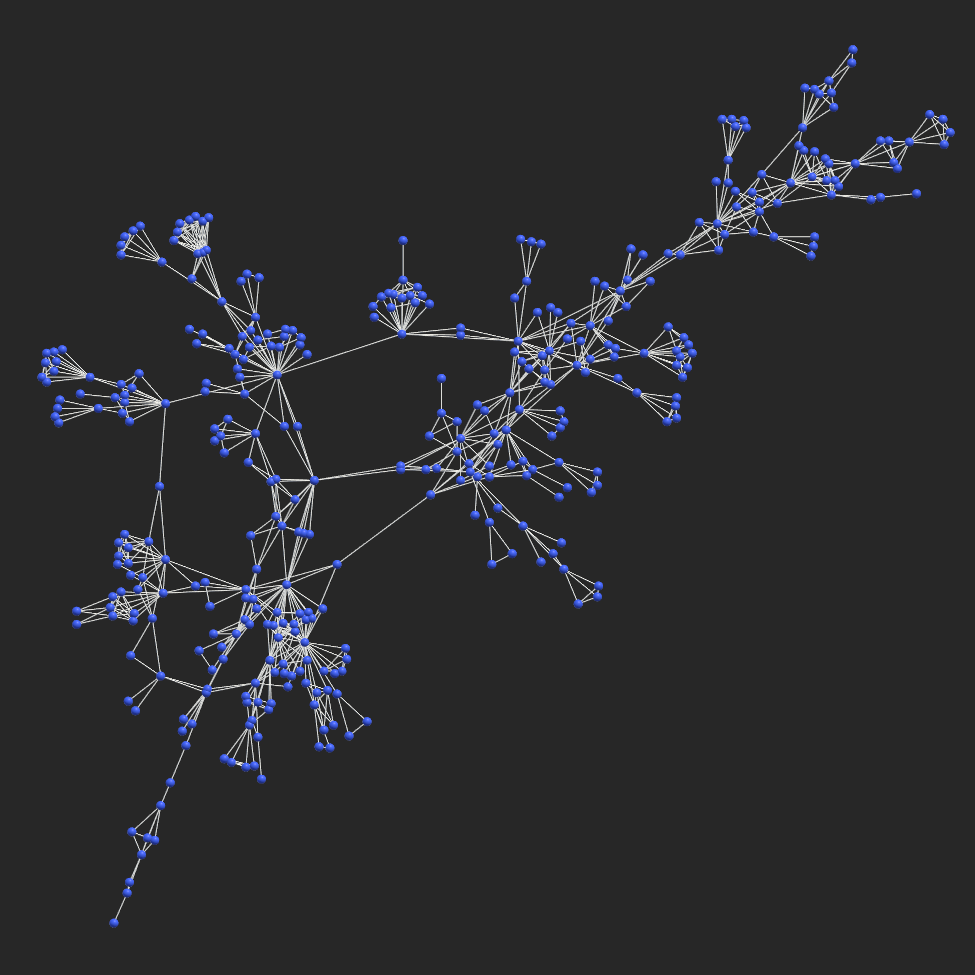}}\hfill
\subfloat[\emph{netsci} (3D)]{\includegraphics[width=0.245\textwidth]{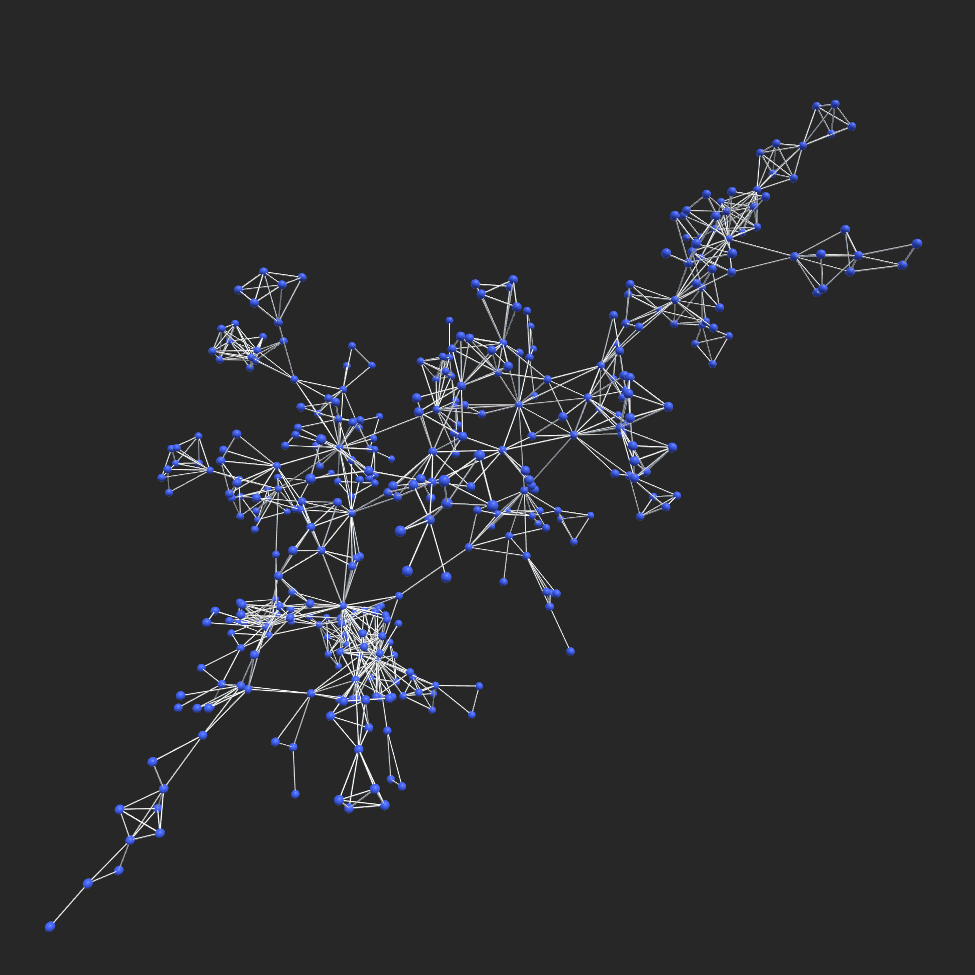}}\hfill\hfill\hfill
\subfloat[\emph{power} (2D)]{\includegraphics[width=0.245\textwidth]{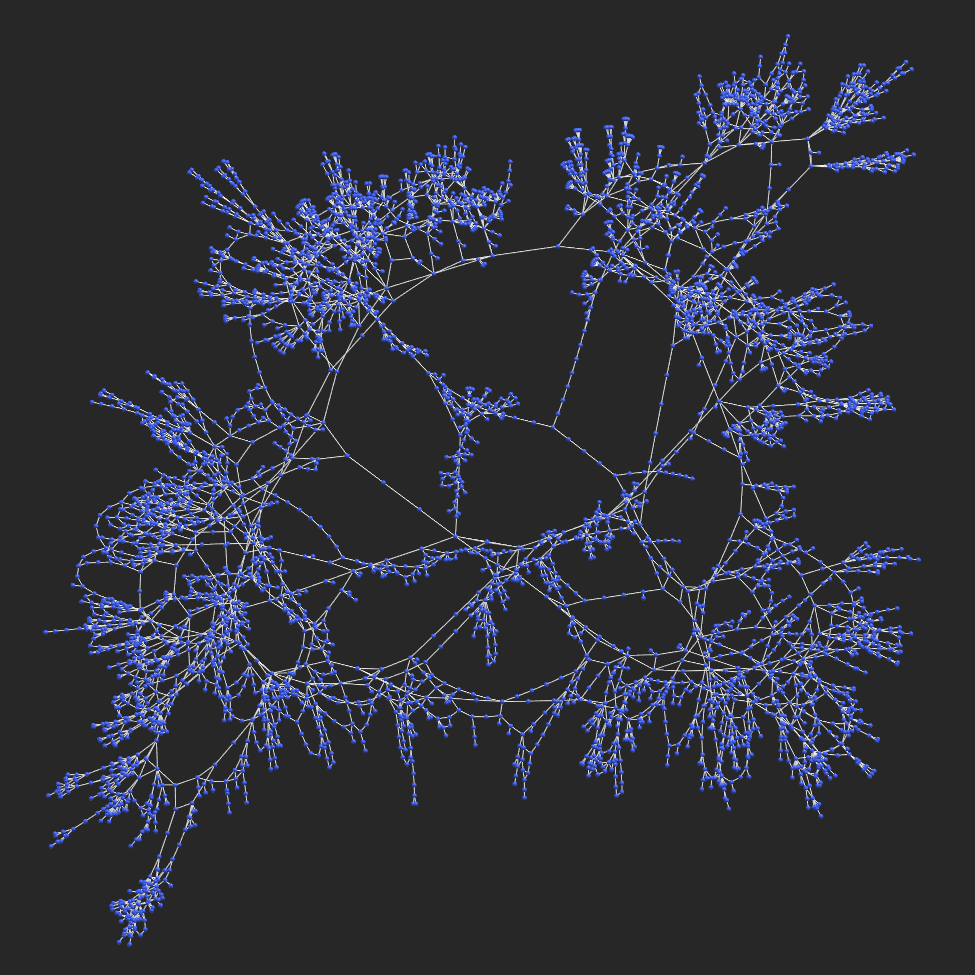}}\hfill
\subfloat[\emph{power} (3D)]{\includegraphics[width=0.245\textwidth]{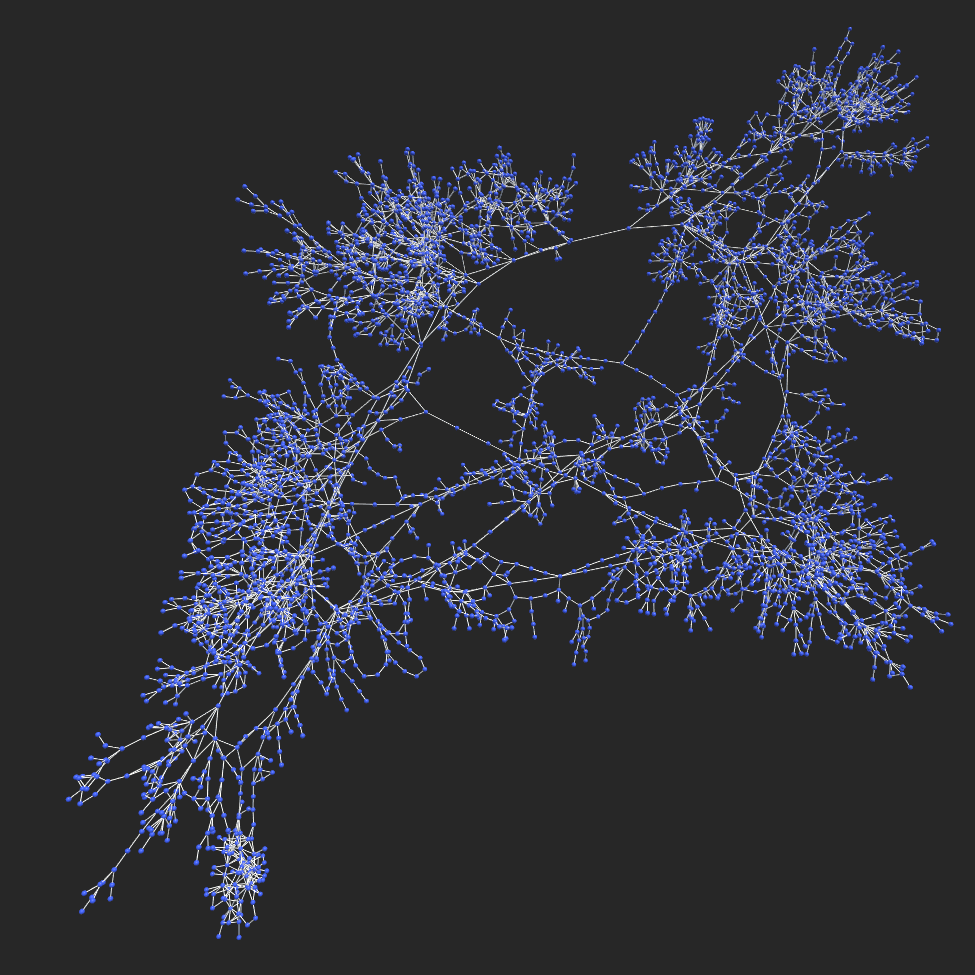}}\\
\caption{The 2D and 3D layouts of the networks used in the experiment.
All layouts are computed using FM$^3$ layout method~\cite{FM3}.
The 2D layouts were shown in a 30-inch desktop display (2,560 $\times$ 1,600 px).
The 3D layouts were rendered in a head-mounted display (2,880 $\times$ 1,600 px, 1,440 $\times$ 1,600 px per eye).
To provide multiple views of a 3D layout from different viewpoints, 
the supplementary material includes videos of the 3D layouts.
Since we used the same layout method (FM$^3$~\cite{FM3}), both 2D and 3D layouts of a network show similar spatial proximity between nodes, while their dimensionalities are different.
This allows a fair comparison between the two visualization conditions (\emph{2D} and \emph{3D}) by reducing the effect of spatial proximity between nodes.
}
\label{fig:layout}
\end{figure*}

The layout of a network is one of the most critical factors in designing a network visualization with a node-link diagram.
While several layout methods are designed specifically for immersive network visualization (e.g., spherical layout methods~\cite{Kwon15, Kwon16}), they often use additional design factors that are outside of the scope of this study (e.g., edge bundling). 
In addition, different layout methods can produce greatly varying layout results~\cite{Gibson12, Kwon19}.
Therefore, we use the same layout method (FM$^3$ layout method~\cite{FM3}) with the same parameter setting for both conditions. 
In both conditions, nodes are drawn as spheres and edges are rendered as thin cylinders with the same color scheme. 
The network visualizations used in the experiment are shown in \autoref{fig:layout} and the supplementary material.
For \emph{2D}, the network is initially laid out to utilize the majority of a 30-inch display, with an aspect ratio of 16:10. 
For \emph{3D}, the network is initially scaled to fill a 1 m$^3$ cube and placed in front of the user in the room-scale area. 

\subsubsection{Interaction}
\label{sec:interaction}
Several interaction techniques are required to carry out the tasks in the experiment.
For example, basic navigation techniques (e.g., zoom and pan) are needed to explore the structure of a network in the visualization.
In addition, users need to select several nodes to complete the tasks (e.g., \autoref{fig:highlight}).

Most users are already familiar with the basic interaction techniques in a desktop environment using mouse and keyboard, such as click-to-select and scroll-to-zoom.
Thus, it is important to design intuitive interaction techniques for the immersive environment to reduce the effect of interaction techniques for completing the tasks, which is not the main focus of this study.

\emph{Selection} in a traditional 2D desktop display with keyboard and mouse is a well-established interaction, which most commonly implemented as a left-click with the mouse. 
However, designing an effective selection interaction for an immersive display environment is not as straightforward as the desktop environment.
Since motion tracking is a common feature of commercially available stereoscopic displays, such as the HTC Vive and Oculus Rift, we leverage the ability to select objects by pointing with a tracked controller.

Users can select a node by first pointing the target node and then pressing a button.
This virtual laser pointer technique is one of the most common ways to implement selection in virtual reality. 
We use this technique because of its intuitiveness and familiarity to most people with any amount of experience in VR or laser pointers.

Our virtual laser pointer implementation does have two minor improvements to make the experience better.
The first is to reduce clutter in the visualization, the laser pointer is only visible while the user touches lightly on the touch-pad of the controller (\autoref{fig:teaser}) and a selection is counted when the user presses the touch-pad until an audible ``click'' is heard.
The other improvement is an error tolerance for selecting distant objects.
This helps significantly, as it is possible with our system to have nodes that take up less than half a degree of the user's field of view, making them very difficult to select with a simple ray casting.
In our pilot study, we found that a cone of approximately two degrees at the apex made selection of distant objects much easier without causing the opposite problem of unintentionally selecting objects far from the pointer.

To match the ease of use to \emph{scale} the network with the mouse scroll wheel in the traditional desktop environment, we implement a pinch-to-zoom technique for the immersive environment. 
To activate the pinch-to-zoom, the user must press a button on both tracked hand-held controllers, then stretch or contract the distance between the two controllers to scale the network by an equivalent factor.
In addition, users can \emph{move} the network by clicking and dragging in the desktop environment, and by dragging the controllers with the pinch-to-zoom button held in the immersive environment.
This design is also motivated by intuitiveness and popularity. 
The pinch-to-zoom technique is widely used in many touch-based systems that most users are already familiar with, such as smartphones and tablets.

In the \emph{2D} condition, \emph{rotating} the network is not necessary for completing the tasks used in this study. 
However, the ability to look at the network from different directions is necessary with the 3D layout to address possible occlusion issues.
Thus, while we exclude the rotation interaction from both conditions, we leverage the physical tracked space allowed by the immersive technology we use in the \emph{3D} condition.
By walking around the space, users can view the network from any angle, even though they cannot rotate the network itself.

\begin{figure*}[t]
\subfloat[\emph{memory} task with the \emph{lesmis} network in the \emph{2D} condition]{\includegraphics[width=0.245\textwidth]{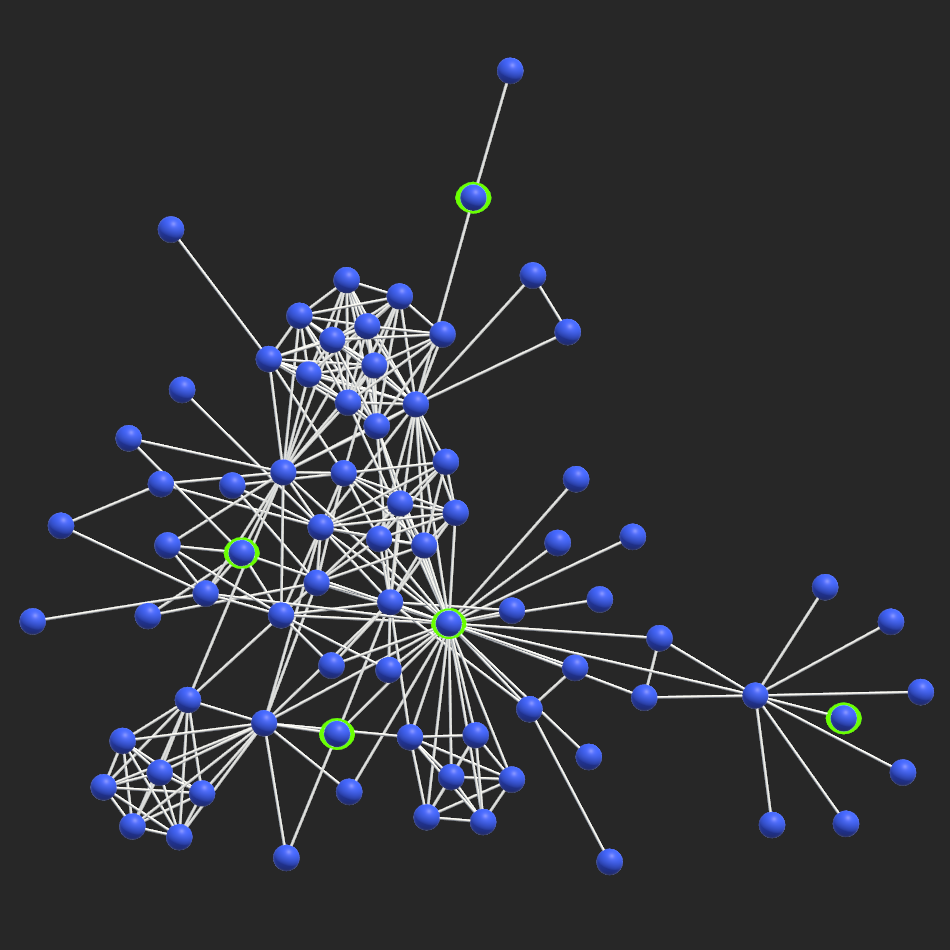}}\hfill
\subfloat[\emph{memory} task with the \emph{lesmis} network in the \emph{3D} condition]{\includegraphics[width=0.245\textwidth]{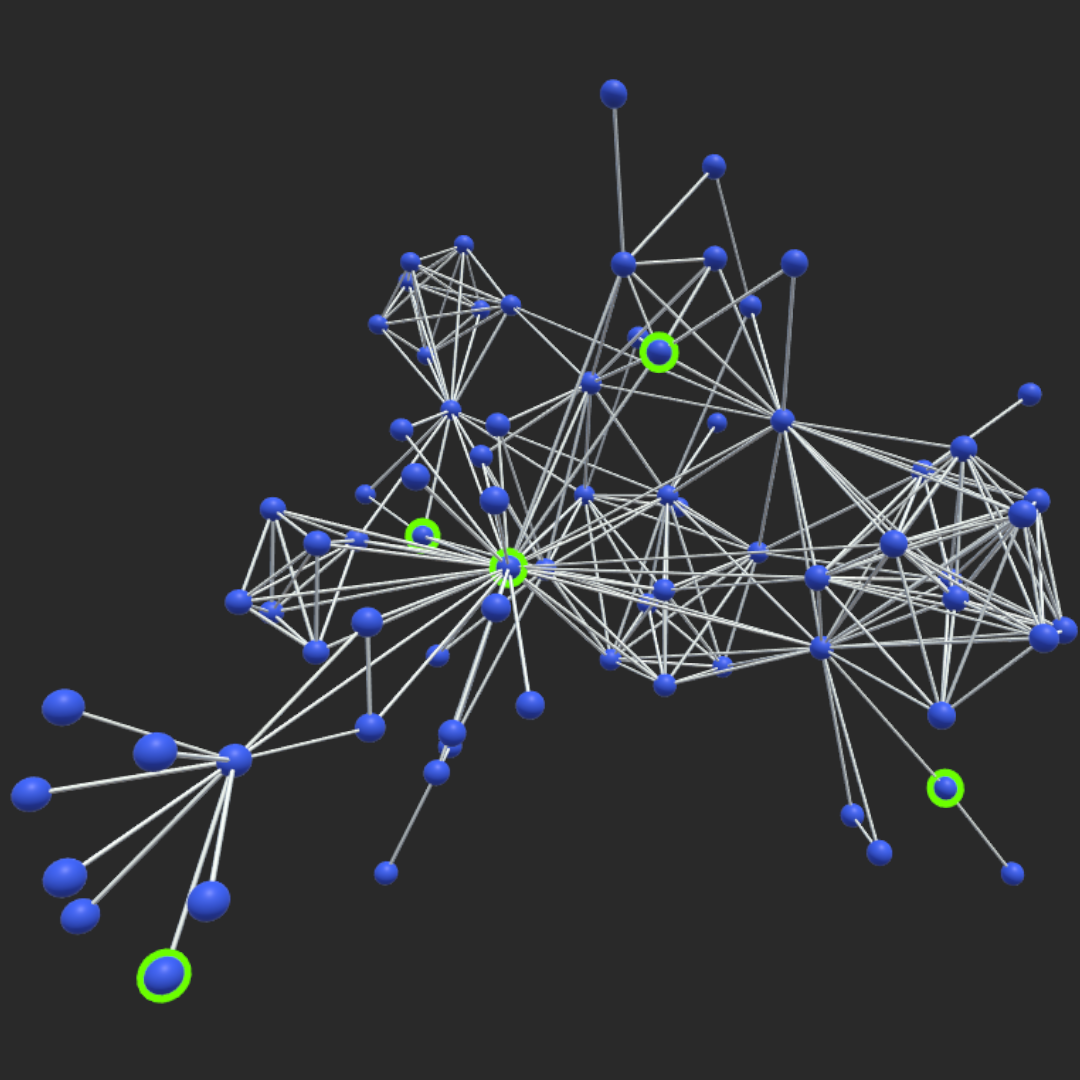}}\hfill\hfill\hfill
\subfloat[\emph{change} task with the \emph{karate} network in the \emph{2D} condition (before change)]{\includegraphics[width=0.245\textwidth]{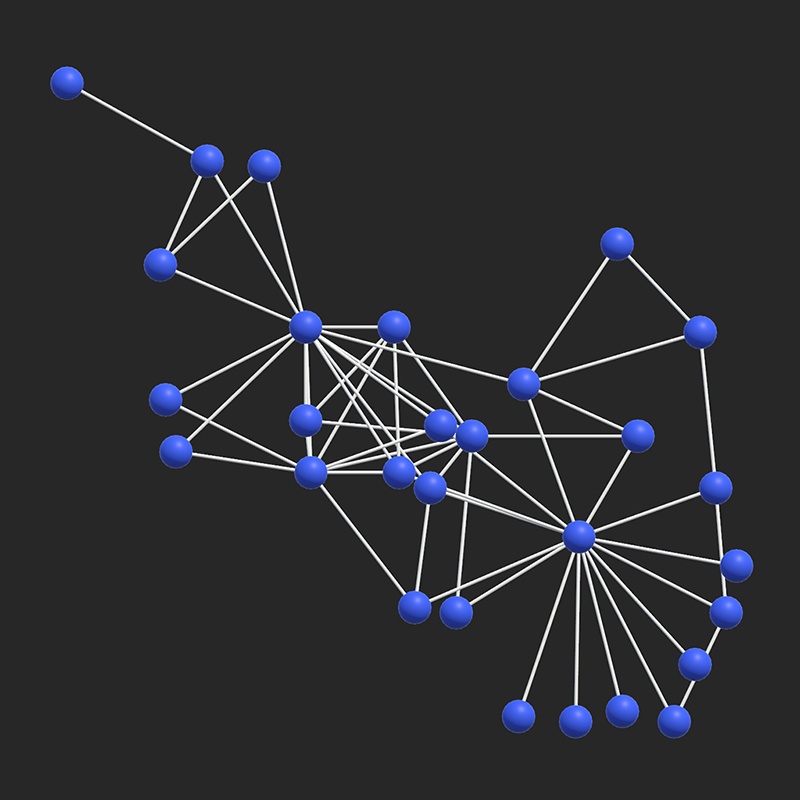}}\hfill
\subfloat[\emph{change} task with the \emph{karate} network in the \emph{2D} condition (after change)]{\includegraphics[width=0.245\textwidth]{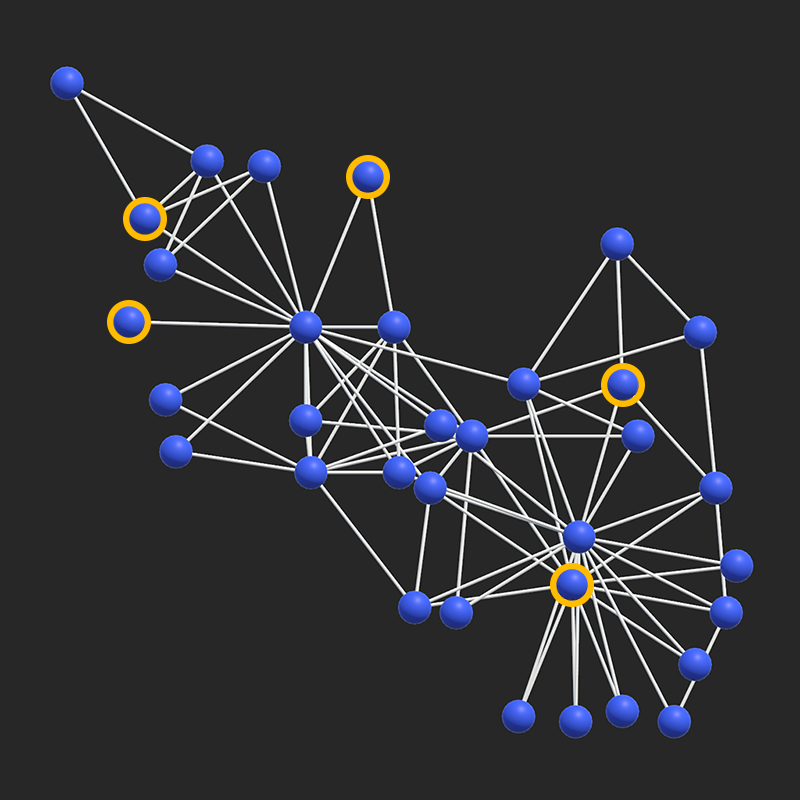}}
\caption{Examples of \emph{memory} and \emph{change} tasks. 
For the \emph{memory} task (a and b), participants explore the visualization for 30 seconds and are asked which nodes were highlighted.
Since the same layout method (FM$^3$~\cite{FM3}) was used for computing both 2D and 3D layouts, the location of target nodes are relatively similar in both 2D and 3D layouts. 
This allows a fair comparison between the two visualization conditions by reducing the effect of the target node location.
For the \emph{change} task, a modified network with five missing nodes (c) is first shown for 30 seconds. 
Then, participants are asked which nodes were missing when shown the full network (d) without any missing nodes.
The highlighted nodes in (d) are the missing nodes in (c).
We used the layout of the full network (d) for the modified network (c) as well since this study focuses on the effect of the two visualization conditions (\emph{2D} and \emph{3D}), not the drawing stability of the layout method we used.
}
\label{fig:recall-and-diff}
\end{figure*}

\subsection{Tasks}
Existing studies have measured the quality of the mental map in the user's mind by human-computer interaction experiments using various tasks.
Archambault and Purchase~\cite{Archambault13} have grouped the tasks into three broad categories: \emph{interpretation}, \emph{memory}, or \emph{change} tasks:
\begin{itemize}[topsep=3pt, partopsep=0pt, itemsep=3pt, parsep=0pt]
    \item An interpretation task asks a question that requires the user to understand the structure information of a network. 
    An example task would be to ask the user about node degree or paths between two nodes in a network. 
    \item A memory task requires the user to recall information about the network after viewing the visualization. For example, reconstructing the network on paper from the user's memory.
    \item A change task asks how the network changes; possible changes are changes in the degree or changes in the overall size.
\end{itemize}
Considering the above task categories by Archambault and Purchase~\cite{Archambault13}, 
we design three tasks for the experiment: 
\begin{itemize}[topsep=3pt, partopsep=0pt, itemsep=3pt, parsep=0pt]
\item \textbf{\emph{Path}}: ``Highlight nodes that form the shortest path from node A to node B.'' 
This is an \emph{interpretation} task~\cite{Archambault13}.
The participant is shown a network with two pre-highlighted nodes.
Participants are asked to find a shortest path between the two pre-highlighted nodes by selecting a set of nodes that forms the shortest path. 
Participants can inspect the connected edges of a node by highlighting the node using the virtual laser pointer (\autoref{fig:highlight} and \autoref{sec:interaction}).
For each network, the same two nodes are pre-highlighted in both visualization conditions. 
While there can be multiple shortest paths, the participant only needs to find one. 

\item \textbf{\emph{Memory}}: ``Select the five nodes that were previously highlighted.''
Participants are given 30 seconds to memorize five pre-highlighted target nodes in a network visualization.
The visualization is then removed, and the participants are shown a blank screen for 10 seconds because immediately showing the network again was found to be too easy in our pilot study.
After that, the network reappears with the same position and scale, but without highlighting the target nodes. 
The goal is to select the five target nodes that were previously highlighted (\autoref{fig:recall-and-diff}a--b).
For each network, the same five nodes are assigned as the target nodes across all visualization conditions and participants.
As we use the same layout method (FM$^3$~\cite{FM3}) for both \emph{2D} and \emph{3D} conditions, the location of target nodes are relatively similar between the two conditions, as shown in \autoref{fig:recall-and-diff}a--b.
This allows a fair comparison between the two conditions by reducing the effect of the target node location.

\item \textbf{\emph{Change}}: ``Select nodes that were not a part of the network.''
Participants are given 30 seconds to explore a modified network, where five target nodes and associated edges are removed from the full network. 
Then, participants are shown a blank screen for 1 second since immediately showing the new network was found to be too easy in our pilot study.
After that, the full network is shown, where the five target nodes and corresponding edges are added back to the network. 
The goal is to identify the five target nodes (\autoref{fig:recall-and-diff}c--d).
We removed the same five nodes from the full network to create the modified network for both visualization conditions and all participants.
We use the layout of the full network for the modified network as well, 
i.e., we did not use a separate layout for the modified network.
This design decision makes our experiment differ from existing experiments on the mental map quality of dynamic network visualization, where different layouts are used per time step of a dynamic network.
However, it allows this study to focus on the effect of the two visualization conditions, not the drawing stability that is outside of the scope of this study.
\end{itemize}
The participants are given a written description of the objective at the start of each task.
The duration of exploration and blank screen stages in the \emph{memory} and \emph{change} tasks are determined based on a pilot study.
For the \emph{path} task, we compute the accuracy of a selected path based on the ratio between the length of the selected path and the length of the shortest path(s).
For the \emph{memory} and \emph{change} tasks, the accuracy of selected nodes by participants is computed based on the graph-theoretic distance from the correct target nodes (more detail in \autoref{sec:result:accuracy}).
\vspace{1em}

\subsection{Networks}
We use three different networks with varying sizes (i.e., number of nodes and edges) with one additional network that was used for the training session. 
\begin{itemize}[topsep=3pt, partopsep=0pt, itemsep=3pt, parsep=0pt]
\item \textbf{\emph{Karate}}: This is the well-known Zachary karate club network~\cite{Karate}, which consists of 34 nodes and 78 edges. This network is used in the training session.
\item \textbf{\emph{Lesmis}}: A co-occurrence network of the characters in Victor Hugo's novel \emph{Les Mis\'{e}rables}~\cite{StandfordGraphBase}, This network has 77 nodes and 254 edges.
\item \textbf{\emph{Netsci}}: A co-authorship network in the field of network science~\cite{Newman06}. We use the largest component of the full network, which consists of 379 nodes and 914 edges. 
\item \textbf{\emph{Power}}: The power grid of western states of the United States~\cite{Watts98}. This network has 4,941 nodes and 6,594 edges.
\end{itemize}

\subsection{Participants} 
We recruited 20 participants (11 female, 9 male) for our user study. The mean age of participants was 24.35, ranging from 18 to 34 years.
Every participant was familiar with the concepts of virtual reality, and 16 participants had used a virtual reality device before this study. 
Of those 16, two participants said they had used virtual reality devices extensively.
Additionally, every participant was familiar with the concept of a network, and 14 said they were familiar or experienced with network data structures.
Four participants were in an undergraduate degree program, while the remaining 16 all possessed undergraduate degrees. 14 of the 16 were also pursuing postgraduate degrees.

Five participants had normal vision, while 15 had corrected vision. 
Of the 15, 13 wore glasses, and all but one were able to wear their glasses comfortably within the HTC Vive Pro HMD.
The one participant that had to remove his/her glasses reported that he/she was still able to read text and see the nodes/edges clearly in the immersive environment.
One participant had deuteranopia vision (self-reported), but reported that all colors used in the study were easily distinguishable.

\subsection{Apparatus}
Both visualization conditions are implemented as Unity3D applications.
For the \emph{3D} condition, participants use the HTC Vive Pro HMD and are instructed to stand in the middle of the room-scale environment at the start of each task, facing toward the area where the networks would be presented. 
The HTC Vive Pro HMD has a 2,880 $\times$ 1,600 px (1,440 $\times$ 1,600 px per eye) AMOLED display with a 90 Hz refresh rate. 
The immersive environment was driven by a desktop computer with an Intel i7 6900K CPU and dual (SLI enabled) NVIDIA GeForce GTX 1080 GPUs.
The environment was consistently rendered at 90 frames per second.
The tracked physical space for room-scale configuration measures 3.0 m by 3.1 m. 
The HTC Vive lighthouses are positioned 3.2 m off the ground. 
The virtual environment is a simple space with a floor indicating the boundary where users can walk.
Participants use a 30-inch desktop display with 2,560 $\times$ 1,600 px for the \emph{2D} condition.

\subsection{Procedure}
After ensuring the participants are aware of possible VR/HMD issues such as sickness or disorientation, we adjust the HMD fit and interpupillary distance (IPD) for each participant to provide optimal viewing conditions. 
Participants were allowed as much time as necessary to ensure the HMD was fit comfortably without issue. 
None of the participants experienced any issue with the HMD.
Participants then completed a pre-study questionnaire, followed by training, the full experiment, and a post-study questionnaire.

\paragraph{Questionnaire.}
All participants answered a two-part questionnaire.
The first part covered participant demographics, including age, gender, education, colorblindness, perceived spatial reasoning skills, and VR/visualization experience levels.
The second part covered perceived task difficulty (7-point Likert scales) as well as free response questions about participant preference.

\paragraph{Training.}
Participants were allowed up to 10 minutes to familiarize themselves with the visualization conditions and their respective interfaces, in a guided tour of the features of both applications.
Each participant received training before performing each task for the first time. 
This training consisted of performing the same set of tasks with the \emph{karate} network, with the correct answer available at the end. 

\paragraph{Procedure.}
The order of the tasks was the same for each participant: \emph{path}, \emph{memory}, and \emph{change} tasks.
The networks were ordered from smallest (\emph{lesmis}) to largest (\emph{power}) within each task.
The order of the visualization conditions was counterbalanced such that half the participants started with \emph{2D} and the other half started with \emph{3D} to prevent learning effects. 
For example, a participant that starts with \emph{2D} would perform tasks in order of \emph{2D-path-lesmis}, \emph{2D-path-netsci}, ..., \emph{2D-memory-lesmis}, \emph{2D-memory-netsci}, ..., then repeating everything with \emph{3D}.

Participants were encouraged to take short rests between tasks, for as long as they needed. 
The visualization was fully reset between tasks. 
For the \emph{3D} condition, the participants were instructed between tasks to re-orient themselves in the middle of the space and face the same direction so that movement in previous tasks does not affect the completion time or success rate of subsequent tasks.
We informed the participants about their right to stop the experiment at any time.

\subsection{Hypotheses} 
We expected the following results from the user study:
\begin{itemize}[topsep=3pt, partopsep=0pt, itemsep=3pt, parsep=0pt]
    \item [H1:] Task completion time will be shorter with traditional display (\emph{2D}) than with immersive display (\emph{3D}).
    \item [H2:] \emph{3D} will outperform \emph{2D} in terms of accuracy.
    \item [H3:] Participants will prefer \emph{3D} over \emph{2D} for exploring network data.
\end{itemize}

\begin{figure*}[t]
\centering
\includegraphics[]{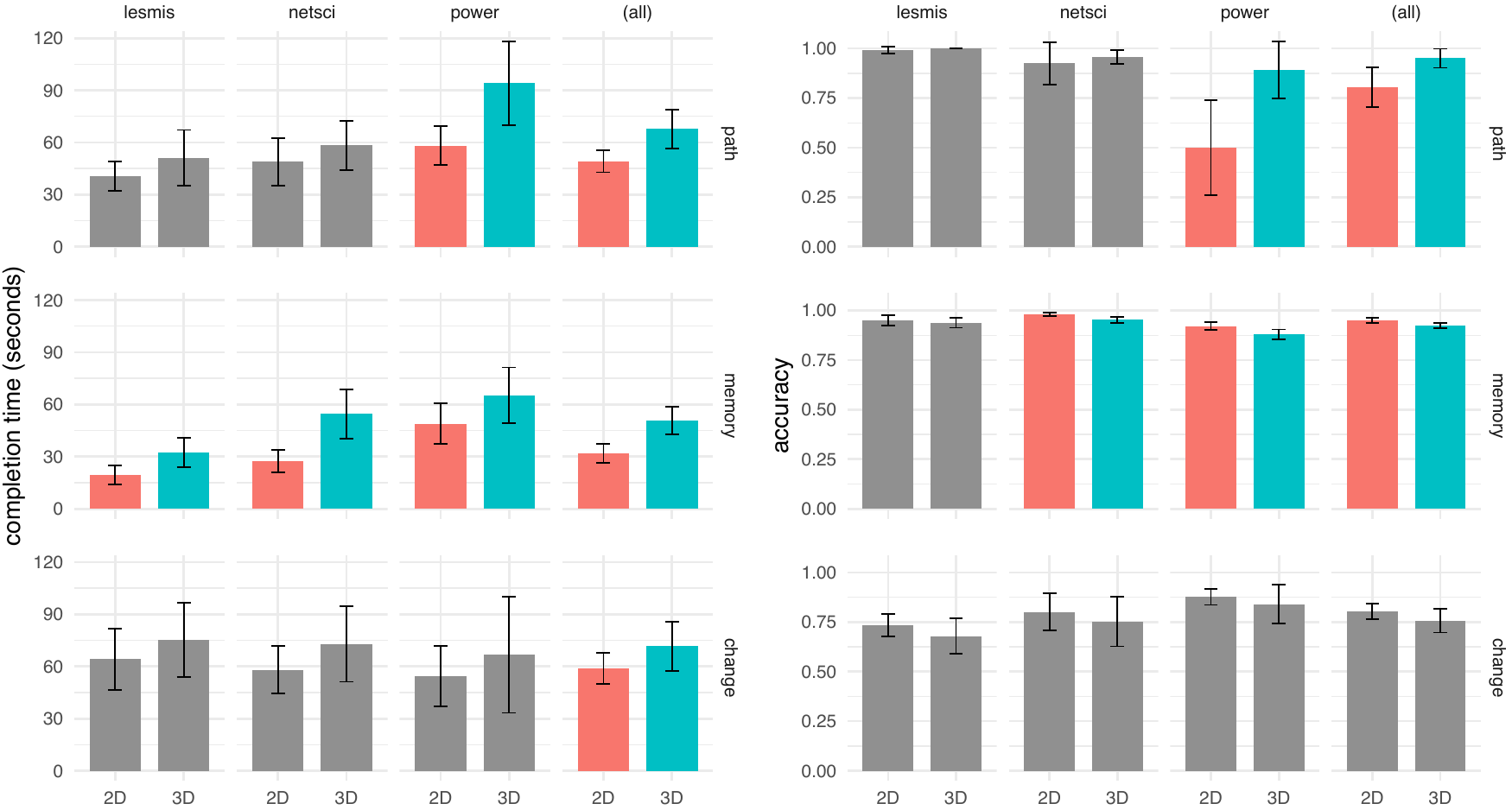}
\label{fig:result}
\vspace{-1em}
\caption{Summary of completion time (left) and accuracy (right) of the experiment. 
Each row of bar charts represents one of the three tasks (\emph{path}, \emph{memory}, and \emph{change}). 
Each column corresponds to one of the three datasets (\emph{lesmis}, \emph{netsci}, and \emph{power}) or all the datasets (\emph{all}). 
Each bar represents the mean value of the corresponding responses.
The error bars represent 95\% confidence intervals.
The colored bars show that the difference between the \emph{2D} and \emph{3D} conditions was statistically significant.
The gray bars mean that there was no statistical significance in the difference between \emph{2D} and \emph{3D}.}
\label{fig:result}
\end{figure*}

%% file: tex/result.tex
\section{Results}
The summary of results are shown in \autoref{fig:result}, where the colored bars show that the difference between \emph{2D} and \emph{3D} was statistically significant for the corresponding task and network. The gray bars in \autoref{fig:result} show that the difference was not statistically significant.
In addition, \autoref{fig:task-difficulty} shows the distribution of task difficulty ratings reported by participants. 

Overall, participants performed faster with the \emph{2D} condition than the \emph{3D} condition for all three tasks, confirming H1.
In terms of accuracy, participants performed better with the \emph{3D} condition than the \emph{2D} condition for the \emph{path} task, as in H2.
However, for the \emph{memory} task, participants' responses were more accurate with the \emph{2D} condition than with the \emph{3D} condition, opposed to H2.
Participants reported that the \emph{path} task was easier in the \emph{3D} condition than the \emph{2D} condition.
An overwhelming majority of participants (85\%) favored the \emph{3D} condition over the \emph{2D} condition, as we expected in H3.

\subsection{Task Completion Time}
Task completion time is measured with a timer built into the system created for this experiment. 
The timer starts when the user is able to start interacting with a network, and it ends when the user pressed the ``continue'' button to proceed to the next task. 
None of the tasks directly indicate that a user has chosen the correct answers, and so some users chose to be more meticulous than others when checking their work. 
As a result, there is high variance in the time required to complete all of the tasks.
We compare each task and also network separately because the completion time varies depending on task and network.

\paragraph{\emph{Path} task.}
On average for both conditions, participants completed the \emph{path} task in 58.53s ($SD=36.51$).
They completed the \emph{path} task faster with the \emph{2D} condition, taking 49.26s ($SD=24.77$), and slower with the \emph{3D} condition, taking 67.81s ($SD=43.60$).
A paired t-test showed that this difference is statistically significant ($t(59) = 3.744$, $p=0.0004$, 95\% CI of the difference $= 8.637$ to $28.467$).
Cohen's $d$ using the pooled standard deviation ($SD_\text{pool}$) suggested that the effect size is medium ($d = 0.523$).

When we analyzed the completion time of \emph{path} task for each network, there was a significant difference only for the \emph{power} network ($t(19) = 3.359$, $p = 0.003$, $95\%$ CI of the difference $= 13.487$ to $58.076$), where participants took 58.22s ($SD=23.79$) with \emph{2D} but took 94.00s ($SD=51.80$) with the \emph{3D} condition. 
Cohen's $d$ using $SD_\text{pool}$ showed that the effect size is large ($d = 0.888$).
We did not observe a significant difference between \emph{2D} and \emph{3D} conditions for \emph{lesmis} ($p = 0.092$) and \emph{netsci} ($p = 0.229$).

\paragraph{\emph{Memory} task.}
On average for both conditions, participants completed the \emph{memory} task in 41.31s ($SD=28.25$).
They completed the \emph{memory} task faster with  the \emph{2D} condition, taking 31.92s ($SD=21.42$), and slower with the \emph{3D} condition, taking 50.70s ($SD=31.18$), which is shown by t-test to be statistically significant ($t(59) = 5.847$, $p = 2.318 \times 10^{-7}$, $95\%$ CI of the difference $= 12.356$ to $25.206$).
Cohen's $d$ using $SD_\text{pool}$ suggested that the effect size is medium to large ($d = 0.702$).

The differences between the \emph{2D} and \emph{3D} conditions for each network were also significant.
For the \emph{lesmis} network, participants completed the \emph{memory} task significantly faster ($t(19) = 4.887$, $p=0.0001$, $95\%$ CI of the difference = 7.339 to 18.333) with the \emph{2D} condition (mean = 19.41s $SD=11.88$) than with the \emph{3D} condition (mean = 32.25s, $SD=18.12$).
Cohen's $d$ using $SD_\text{pool}$ showed that the effect size is large ($d = 0.838$).
For the \emph{netsci} network, participants completed the \emph{memory} task significantly faster with the \emph{2D} condition than with the \emph{3D} condition, having mean completion times of 27.38s ($SD=13.54$) and 54.59s ($SD=30.30$), respectively ($t(19) = 3.959$, $p=0.0008$, $95\%$ CI of the difference $= 12.824$ to $41.596$).
Cohen's $d$ using $SD_\text{pool}$ showed that the effect size is large ($d = 1.160$).
Participants also completed the \emph{memory} task significantly faster ($t(19) = 2.714$, $p = 0.014$, $95\%$ CI of the difference $= 3.729$ to $28.856$) with the \emph{2D} condition than with the \emph{3D} condition for the \emph{power} network, having mean completion times of 48.96s ($SD=24.65$) and 65.26s ($SD=34.26$), respectively.
Cohen's $d$ using $SD_\text{pool}$ suggested that the effect size is medium ($d = 0.546$).

\paragraph{\emph{Change} task.}
On average for both conditions, participants completed the \emph{change} task in 65.28s ($SD=46.16$).
They completed the \emph{change} task faster with the \emph{2D} condition (mean = 58.86s $SD=34.54$) than with the \emph{3D} condition (mean = 71.69, $SD=54.97$).
A paired t-test shows that this difference is statistically significant ($t(59) = 2.397$, $p=0.020$, 95\% CI of the difference $= 2.118$ to $23.546$.
However, Cohen's $d$ using $SD_\text{pool}$ showed that the effect size is small ($d = 0.280$).
In addition, there was no significant difference between the \emph{2D} and \emph{3D} conditions when we analyzed the completion time of the \emph{change} task for each network individually.

\subsection{Accuracy} 
\label{sec:result:accuracy}
We define the accuracy of a response according to each task. 

\paragraph{\emph{Path} task.}
For the \emph{path} task, the accuracy (\autoref{eq:accuracy-shortest-path}) is based on the ratio between the length of the selected path and the length of the shortest path(s).
The participants were asked to find the shortest path between two pre-highlighted nodes, and so the correct answer would be any complete path that is the same length as the shortest path between those two nodes. 
Thus, we use the shortest path length as a fraction of the participant's selected path when the participant makes a valid path, and an accuracy of zero is used for invalid paths (e.g., disconnected nodes), shown in the following equation. 
\begin{equation}
\text{accuracy} = \frac{\text{the length of the shortest path}}{\text{the length of the selected path}}
\label{eq:accuracy-shortest-path}
\end{equation}
With this definition, the best accuracy score a participant can receive is 1, and any path that is valid but contains extra nodes will be between 0 and 1, approaching 0 as more redundant nodes are added.

Overall, participants succeeded in making a valid path in 89.17\% of all responses, and that path was shortest in 80.83\% of all responses. 
Participants performed significantly better in the \emph{3D} condition with the mean accuracy of 0.950 out of 1.0 ($SD=0.185$), compared to the mean accuracy in the \emph{2D} condition, at 0.806 ($SD=0.387$). 
A paired t-test confirms that this difference is statistically significant ($t(59) = 2.820$, $p=0.0065$, $95\%$ CI of the difference $= 0.042$ to $0.248$).
Cohen's $d$ using $SD_\text{pool}$ suggested that the effect size is small to medium ($d = 0.478$).

When we analyzed the accuracy for the \emph{path} task per network separately, there was a significant difference only for the \emph{power} network ($t(19) = 2.970$, $p = 0.008$, $95\%$ CI of the difference $= 0.116$ to $0.670$), where the mean accuracy was 0.500 ($SD=0.513$) with \emph{2D} but was 0.893 ($SD=0.307$) with the \emph{3D} condition. 
Cohen's $d$ using $SD_\text{pool}$ suggested that the effect size is small to large ($d = 0.929$).
We did not observe a significant difference between \emph{2D} and \emph{3D} conditions for \emph{lesmis} ($p = 0.330$) and \emph{netsci} ($p = 0.479$).

\paragraph{\emph{Memory} task.}
If we only consider whether a selected node is a target node or not, it is not possible to measure how close the selected node is to a target node.
Thus, we define the accuracy of a response for the \emph{memory} task based on the graph-theoretic distance between a selected node by participants and the closest target node:
\begin{equation}
    \text{accuracy} = 1 - \frac{\text{the distance from the selected node}}{\text{the diameter of the network}}
\label{eq:accuracy-memory-change}
\end{equation}
We normalize the distance by the diameter of a network, which is the maximum distance between any pair of nodes in the network.
Using this definition, the accuracy of a correctly selected node is~1, and the accuracy of an incorrectly selected node will be between 0 and 1.
We use the mean of this accuracy score of participant's responses.
Some participants provided fewer than five responses (i.e., the number of target nodes). 
In that case, we considered the accuracy of the remaining responses is 0 as they could not remember some target nodes at all.
For example, if a participant exactly identified three target nodes (accuracy of 1) but did not provide any more responses (accuracy of 0), the accuracy of the responses by this participant is~0.6.
A single target node can be the closest one to multiple selected nodes, which can lead to incorrect accuracy scores. 
However, there were no responses that had this issue.

Although Euclidean distance is a possible option for measuring the accuracy of a response instead of graph-theoretic distance, the physical sizes of the \emph{2D} and \emph{3D} conditions are quite different. 
Also, due to different dimensionalities of 2D and 3D layouts of a network, it is not clear to define a normalization scheme for comparing Euclidean distances in the two visualization conditions.
Thus, we use the graph-theoretic distance for measuring the accuracy of response.

For all conditions and networks, the mean accuracy of responses was 0.937 ($SD=0.053$) for the \emph{memory} tasks, where participants exactly identified 62.67\% of the target nodes.
The accuracy of responses were higher in the \emph{2D} condition (mean $= 0.951$, $SD = 0.048$) than the \emph{3D} condition (mean $= 0.923$, $SD = 0.056$).
A paired t-test revealed that the difference is statistically significant ($t(59) = 4.2086$, $p = 8.88 \times 10^{-5}$, $95\%$ CI of the difference $= 0.014$ to $0.040$).
Cohen's $d$ using $SD_\text{pool}$ showed that the effect size is medium ($d = 0.527$).

Comparing \emph{memory} task accuracy per network separately shows that the differences in mean accuracy between the \emph{2D} and \emph{3D} conditions were significant for the \emph{netsci} and \emph{power} networks, but not for the \emph{lesmis} network ($p = 0.316$).
For the \emph{netsci} network, the mean accuracy of responses with the \emph{2D} condition was 0.981 ($SD = 0.017$) while the mean accuracy with \emph{3D} was 0.952 ($SD = 0.031$), and the difference was statistically significant confirmed by a paired t-test ($t(19) = 3.940$, $p = 0.0009$, $95\%$ CI of the difference $= 0.014$ to $0.044$).
Cohen's $d$ using $SD_\text{pool}$ showed that the effect size is large ($d = 1.146$).
For the \emph{power} network, a paired t-test showed that the difference of the mean accuracy between \emph{2D} (mean $= 0.921$, $SD = 0.045$) and \emph{3D} (mean $= 0.880$, $SD = 0.052$) conditions was statistically significant ($t(19) = 3.058$, $p = 0.0065$, $95\%$ CI of the difference $= 0.013$ to $0.069$).
Cohen's $d$ using $SD_\text{pool}$ suggested that the effect size is large ($d = 0.852$).

\paragraph{\emph{Change} task.}
We use the accuracy definition of the \emph{memory} task (\autoref{eq:accuracy-memory-change}) for the \emph{change} task as well.
Overall, the mean accuracy for all networks and conditions were 0.780 ($SD = 0.196$).
The mean accuracy was higher in the \emph{2D} condition (mean $= 0.804$, $SD = 0.153$) than the \emph{3D} condition (mean $= 0.756$, $SD = 0.231$). 
However, the difference was not statistically significant ($p = 0.458$).
Moreover, there were no significant differences between \emph{2D} and \emph{3D} conditions when we analyzed for each network individually.

\begin{figure}[t]
\centering
\includegraphics[]{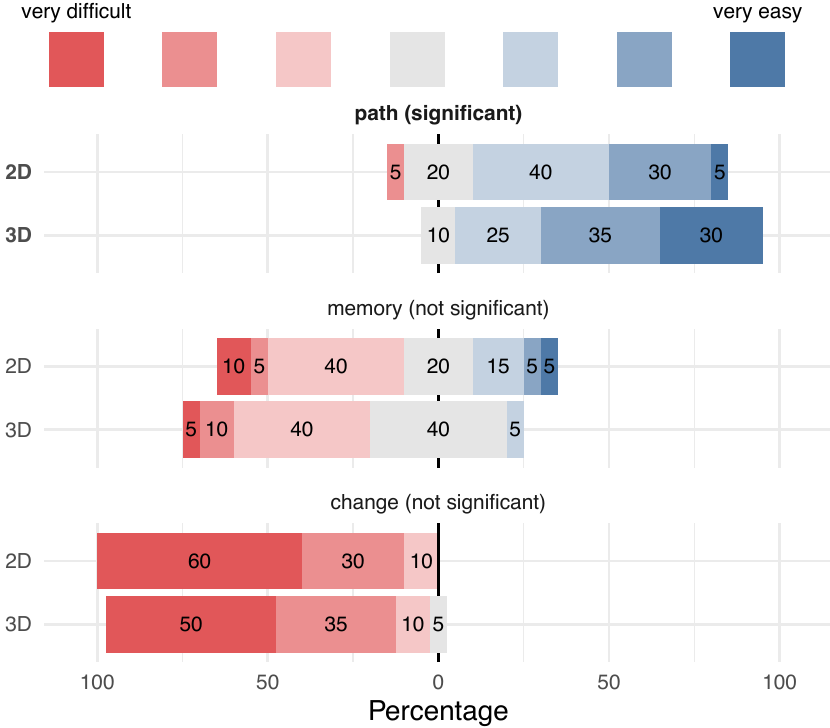}
\caption{User responses to questions about the difficulty of each task and visualization condition.}
\label{fig:task-difficulty}
\end{figure}

\subsection{User Feedback}
Participants were asked to complete post-study questionnaires that provide more insight into what each person experienced between the two visualization conditions.

We asked participants to rate the perceived difficulty of each task with each condition on a 7-point Likert scale, where 1 meant ``very difficult'' and 7 meant ``very easy.''
The overall distribution of the responses is shown in \autoref{fig:task-difficulty}.
Participants reported that the \emph{path} task was easier using the \emph{3D} condition (median $= 6$, IQR $= 2$) than \emph{2D} (median $= 5$, IQR $= 1.25$).
A Wilcoxon signed-rank test (a non-parametric test equivalent to the paired t-test) showed that the difference was statistically significant ($V = 11$, $p = 0.007$).
However, participants reported that the \emph{2D} and \emph{3D} conditions are not different in terms of difficult for \emph{memory} ($V = 78$, $p = 0.306$) and \emph{change} ($V = 25$, $p = 0.478$) tasks.

It is clear, however, that the \emph{path} task was the easiest task (median~$=~5.5$, IQR~$=~1$), the \emph{change} task was the hardest (median~$=~1$, IQR~$=~1$), while the \emph{memory} task was somewhere in between (median~$=~3$, IQR~$=~1$).
A Friedman test (non-parametric alternative to one-way repeated measures ANOVA) showed that the difference is statistically significant ($\chi^2(2)=71.563$,  $p~=~2.886 \times 10^{-16}$)

85\% of the participants (17 out of 20) reported that they preferred \emph{3D} over \emph{2D}, showing a clear majority.
To the question of what they liked about either \emph{2D} or \emph{3D}, the following lists some of the participants' most common and insightful responses, starting with most the frequent sentiments:
\begin{itemize}[topsep=3pt, partopsep=0pt, itemsep=3pt, parsep=0pt]
    \item \emph{``3D makes it easier to grasp the structure.''}
    \item \emph{``3D networks are easier and/or more interesting to explore''}
    \item \emph{``Edge occlusion was much easier to resolve within 3D than 2D space.''}
    \item \emph{``Node selection is too much effort in 3D}
    \item \emph{``2D layout is better for quick overview and memory''}
\end{itemize}

%% file: tex/discussion.tex
\section{Discussion}
The results suggest that one visualization condition did not consistently outperform the other. 
Although participants took more time to complete tasks with the immersive 3D visualization, they interpreted the structure of a network more accurately (\emph{path} task).
The traditional 2D desktop visualization environment was better for the \emph{memory} task, where participants performed better in terms of both task completion time and accuracy than the immersive 3D condition.
In addition, the two visualization conditions performed similarity in the \emph{change} task except for the overall task completion time.

Finding the shortest path was significantly easier and more accurate with the immersive 3D visualization than the traditional 2D desktop environment.
A significant portion of this can be attributed to the nature of 3D layouts having much fewer \emph{edge crossings}. 
Since the 3D layouts used in the immersive space were almost completely devoid of any crossed edges, it was never the case that a user would be unable to tell whether the path they had highlighted was complete or not, as each edge was easy to follow through the environment from start to finish. 
Even with unavoidable occlusions in 3D, participants could  separate the edges by simply moving their head to change the viewpoint.
Several participants commented on this:
\begin{itemize}[topsep=3pt, partopsep=0pt, itemsep=3pt, parsep=0pt]
\item \emph{``Easy to see the links between the nodes (3D)''}
\item \emph{``Better sense of the topological structure (3D)''}
\item \emph{``3D makes it easier to grasp more of the structure of the graph''}
\item \emph{``It looked like they overlapped and it was harder to tell where the nodes were on the 2D display.''}
\item \emph{``It felt easier to see and understand the structure of the graph in 3D.''}
\item \emph{``VR version takes advantage of depth to make graph clearer.''}
\item \emph{``Edge overlapping was a lot easier to resolve in VR than in 2D space''}
\end{itemize}
In the case of the traditional 2D environment, and especially with the largest network (\emph{power}), it was common for edges to not only cross other edges, but also travel through nodes, making it unclear whether it was a single edge that went through the node, or two edges that happened to connect to the same node at opposite sides. 
If not for the edge-highlighting that was built into the system, it would be obvious that a 3D layout should do much better and any other factors would likely be insignificant. However, there is likely more at play here. 
The edge-highlighting allowed participants to see very precisely if two nodes were connected simply by hovering over it, and all participants were instructed in the use of this feature. 
In spite of this, participants with the \emph{2D} condition still marked incomplete paths and complete but non-optimal paths more frequently than participants with the \emph{3D} condition. 
In short, the immersive experience helped participants complete the path finding task more accurately beyond the 3D nature of the layout. 
This could be evidence of the natural formation of a 3D mental map that helped participants understand the structure of a network and connections between nodes better than could happen in the 2D traditional display environment.

Participants performed slower with the immersive 3D environment in most regards compared to the traditional 2D setup. 
Part of this may have been due to participants' lack of familiarity with the controls and virtual environment. 
In addition, the ability to show a better overview of the \emph{2D} condition seems to affect the overall completion time. 
Some comments from participants are related to these findings:
\begin{itemize}[topsep=3pt, partopsep=0pt, itemsep=3pt, parsep=0pt]
    \item \emph{``Selecting a node is much easier in 2D''}
    \item \emph{``Selection of nodes was easier with a mouse than the immersive 3D controller due to a mouse being more precise and not prone to unintended movements.''}
    \item \emph{``For very precise applications, I might find it useful to use the VR display in conjunction with traditional mouse+keyboard controls.''}
    \item \emph{``The desktop gave me more familiarity in terms of viewing the graph and how to maneuver it.''}
    \item \emph{``A graph can fit within a window size well in 2D''}
    \item \emph{``2D is easy to remember the positions''}
    \item \emph{``It is easier to have an overview with the desktop''}
\end{itemize}
Since the \emph{2D} condition did not require any rotation, we also exclude the rotation interaction in the \emph{3D} condition to remove a possible confounding factor.
In addition, many participants in the pilot study were disoriented after a few rotations of the network.
Thus, we let users view the network from different angles by walking around the space rather than directly rotating the network.
However, the observations on participants' movement during the experiment and the feedback from participants suggest that the physical movements required for finding good viewpoints might lead to taking more time in the immersive environment. 
Two participants specifically commented that rotating the network visualization with the controllers will be helpful:
\begin{itemize}[topsep=3pt, partopsep=0pt, itemsep=3pt, parsep=0pt]
    \item \emph{``I felt that having the ability to rotate the graph in VR would have been helpful.''}
    \item \emph{``Add rotation to VR especially.''}
\end{itemize}
However, these two participants reported that they are moderately or highly experienced with virtual reality technologies.
Investigating the effect of spatial awareness in immersive visualization can be an interesting research topic for future study.

In addition, participants in the immersive environment took more time not necessarily because they didn't find answer as quickly, but because they often spent extra time exploring the network or having fun with the visualization before continuing to the next task, something they did not do with the traditional visualization conditions. 
Here are several related comments from participants:
\begin{itemize}[topsep=3pt, partopsep=0pt, itemsep=3pt, parsep=0pt]
    \item \emph{``It's fun to zoom in and zoom out in VR settings''}
    \item \emph{``3D was more fun and it is easier to visualize where the nodes were.''}
    \item \emph{``I liked the 3D version because it allowed me to step inside the graph and see the nodes hidden in the back.''}
    \item \emph{``3D makes it easier to explore.''}
\end{itemize}
The mental maps formed through this experiment seem to be limited due to the short time frame of all the tasks. 
Since participants were given only 30 seconds to scrutinize each visualization (\emph{memory} and \emph{change} tasks), they only had time to construct a relatively immature mental map. 
This mental map was too fragile for immersive space, based on the overall lower accuracy in immersive 3D for the \emph{memory} task. 
Participants mentioned on several occasions that after having changed their perspective in the immersive environment, they lost their track of the locations of target nodes in the \emph{memory} task.
The difficulty of the \emph{change} task seems to have contributed to some unexpected results. 
Participants answered overwhelmingly that this task was the hardest. 
While all participants finished the entire set of trials, 
a few participants gave up on completing the task for the largest network (\emph{power}), by either selecting randomly or selecting less than five nodes.
We believe the results can be different if more time was allowed for explore and memorize a network for both \emph{memory} and \emph{change} tasks.

Another way to reduce the difficulty of the \emph{change} task is  allowing users to see both structures (before and after adding nodes), similar to the experiments of drawing stability in dynamic network visualization techniques.
However, this could make the task too easy, which is what we found during our pilot study. 
In dynamic network visualization, different layouts are used for each time step, while we used a single layout for both steps as our focus was not the drawing stability.
This leads to another research question: would immersive environments be helpful for dynamic network visualization?

To achieve high internal validity, we have carefully designed the two visualization conditions by (1) keeping the factors that are not the focus of the study the same in both conditions (i.e., the same layout algorithm, interaction paradigm, and color scheme) and (2) counterbalancing the order of visualization conditions.
In addition, the user's performance is measured using well-defined tasks and measurements to have high construct validity.

However, our study has limitations in terms of external validity (generalizability).
Since we considered the entire visualization environment as a single condition, our study only shows the effect of the combination of multiple factors of each environment, but not the effect of individual factors.
Thus, the results can vary, for example, with different display and interaction devices.
Moreover, different data and tasks can lead to different results.
Further studies are needed to evaluate various factors of network visualization.

%% file: tex/conclusion.tex
\section{Conclusion}
Literature on the mental map in network visualization is substantial but sparsely populated when it comes to the immersive visualization design, despite the rapid development of immersive technologies.  
This work investigates the mental maps of viewers under different visualization conditions.
The immersive 3D network visualization proved to be more helpful for interpreting the structure of a network.
It suggests that immersive technology may allow deeper and more intuitive understanding of abstract data structures, though it is more carefully considered and must be precisely designed.
The traditional 2D visualization has proven superior for spatial memory as they provide a better overview of the entire network.

To further investigate the use of immersive technology in understanding abstract data structures, such as networks, the next step is to delve deeper into the individual processes for observing and understanding immersive environments with longer and more detailed experiments. 
These are processes such as how users mentally store relative positions and directions, and how they are able to turn that into an understanding of the data presented. 
Investigating the mental capacity of humans in immersive environments will help us design better systems that leverage human perception more effectively.

%% file: main_pvis.bbl
\begin{thebibliography}{10}

\bibitem{Archambault18Shonan}
D.~Archambault, K.~Klein, and K.~Misue.
\newblock {Reimagining the Mental Map and Drawing Stability}.
\newblock Technical Report 2018-12, National Institute of Informatics Shonan
  Meeting, 2018.

\bibitem{Archambault11}
D.~Archambault, H.~Purchase, and B.~Pinaud.
\newblock {Animation, Small Multiples, and the Effect of Mental Map
  Preservation in Dynamic Graphs}.
\newblock {\em IEEE Transactions on visualization and Computer Graphics},
  17(4):539--552, 2011.

\bibitem{Archambault12}
D.~Archambault and H.~C. Purchase.
\newblock {Mental Map Preservation Helps User Orientation in Dynamic Graphs}.
\newblock In {\em International Symposium on Graph Drawing}, pages 475--486.
  Springer, 2012.

\bibitem{Archambault13}
D.~Archambault and H.~C. Purchase.
\newblock {The ``Map'' in the Mental Map: Experimental Results in Dynamic Graph
  Drawing}.
\newblock {\em International Journal of Human-Computer Studies},
  71(11):1044--1055, 2013.

\bibitem{Ball05}
R.~Ball and C.~North.
\newblock {Effects of Tiled High-Resolution Display on Basic Visualization and
  Navigation Tasks}.
\newblock In {\em Proc. ACM CHI '05 Extended Abstracts on Human Factors in
  Computing Systems}, pages 1196--1199, 2005.

\bibitem{Bryson96}
S.~Bryson.
\newblock {Virtual Reality in Scientific Visualization}.
\newblock {\em Commun. ACM}, 39(5):62--71, 1996.

\bibitem{Buschel18}
W.~B{\"u}schel, J.~Chen, R.~Dachselt, S.~Drucker, T.~Dwyer, C.~G{\"o}rg,
  T.~Isenberg, A.~Kerren, C.~North, and W.~Stuerzlinger.
\newblock {Interaction for Immersive Analytics}.
\newblock In {\em Immersive Analytics}, pages 95--138. Springer, 2018.

\bibitem{Cordeil17}
M.~Cordeil, T.~Dwyer, K.~Klein, B.~Laha, K.~Marriott, and B.~H. Thomas.
\newblock {Immersive Collaborative Analysis of Network Connectivity: CAVE-style
  or Head-Mounted Display?}
\newblock {\em IEEE Transactions on visualization and Computer Graphics},
  23(1):441--450, 2017.

\bibitem{DeRidder15}
M.~de~Ridder, Y.~Jung, R.~Huang, J.~Kim, and D.~D. Feng.
\newblock {Exploration of Virtual and Augmented Reality for Visual Analytics
  and 3D Volume Rendering of Functional Magnetic Resonance Imaging (fMRI)
  Data}.
\newblock In {\em Proc. International Symposium on Big Data Visual Analytics},
  pages 1--8. IEEE, 2015.

\bibitem{Diehl01}
S.~Diehl, C.~G{\"o}rg, and A.~Kerren.
\newblock {Preserving the Mental Map using Foresighted Layout}.
\newblock In D.~S. Ebert, J.~M. Favre, and R.~Peikert, editors, {\em Data
  Visualization 2001}, pages 175--184, Vienna, 2001. Springer Vienna.

\bibitem{Drogemuller18}
A.~Drogemuller, A.~Cunningham, J.~Walsh, M.~Cordeil, W.~Ross, and B.~Thomas.
\newblock {Evaluating Navigation Techniques for 3D Graph Visualizations in
  Virtual Reality}.
\newblock In {\em Proc. International Symposium on Big Data Visual and
  Immersive Analytics}, pages 1--10. IEEE, 2018.

\bibitem{Dwyer18}
T.~Dwyer, K.~Marriott, T.~Isenberg, K.~Klein, N.~Riche, F.~Schreiber,
  W.~Stuerzlinger, and B.~H. Thomas.
\newblock {Immersive Analytics: An Introduction}.
\newblock In {\em Immersive Analytics}, pages 1--23. Springer, 2018.

\bibitem{cave2}
A.~Febretti, A.~Nishimoto, T.~Thigpen, J.~Talandis, L.~Long, J.~D. Pirtle,
  T.~Peterka, A.~Verlo, M.~Brown, D.~Plepys, D.~Sandin, L.~Renambot,
  A.~Johnson, and J.~Leigh.
\newblock {CAVE2: A Hybrid Reality Environment for Immersive Simulation and
  Information Analysis}.
\newblock In {\em Proc. Engineering Reality of Virtual Reality}, 2013.

\bibitem{Gibson12}
H.~Gibson, J.~Faith, and P.~Vickers.
\newblock {A Survey of Two-dimensional Graph Layout Techniques for Information
  Visualization}.
\newblock {\em Information Visualization}, 12(3--4):324--357, 2013.

\bibitem{Greffard12}
N.~Greffard, F.~Picarougne, and P.~Kuntz.
\newblock Immersive dynamic visualization of interactions in a social network.
\newblock In {\em Challenges at the Interface of Data Analysis, Computer
  Science, and Optimization}, pages 255--262. Springer, 2012.

\bibitem{FM3}
S.~Hachul and M.~J{\"u}nger.
\newblock {Drawing Large Graphs with a Potential-Field-Based Multilevel
  Algorithm}.
\newblock In {\em Proc. Graph Drawing}, pages 285--295, 2004.

\bibitem{Hall98}
C.~R. {Hall}, R.~J. {Stiles}, and C.~D. {Horwitz}.
\newblock {Virtual Reality for Training: Evaluating Knowledge Retention}.
\newblock In {\em Proc. IEEE Virtual Reality Annual International Symposium},
  pages 184--189, 1998.

\bibitem{Herman00}
I.~Herman, G.~Melan\c{c}on, and M.~S. Marshall.
\newblock {Graph Visualization and Navigation in Information Visualization: A
  Survey}.
\newblock {\em IEEE Transactions on visualization and Computer Graphics},
  6(1):24--43, 2000.

\bibitem{Huang17}
Y.-J. Huang, T.~Fujiwara, Y.-X. Lin, W.-C. Lin, and K.-L. Ma.
\newblock {A Gesture System for Graph Visualization in Virtual Reality
  Environments}.
\newblock In {\em Proc. IEEE Pacific Visualization Symposium}, pages 41--45.
  IEEE, 2017.

\bibitem{StandfordGraphBase}
D.~E. Knuth.
\newblock {\em {The Stanford GraphBase: A Platform for Combinatorial
  Computing}}.
\newblock Addison-Wesley, 1993.

\bibitem{Krokos18}
E.~Krokos, C.~Plaisant, and A.~Varshney.
\newblock {Virtual Memory Palaces: Immersion Aids Recall}.
\newblock {\em Virtual Reality}, pages 1--15, 2018.

\bibitem{Kwon19}
O.-H. Kwon and K.-L. Ma.
\newblock {A Deep Generative Model for Graph Layout}.
\newblock {\em IEEE Transactions on visualization and Computer Graphics},
  26(1):665--675, 2020.

\bibitem{Kwon15}
O.-H. Kwon, C.~Muelder, K.~Lee, and K.-L. Ma.
\newblock {Spherical Layout and Rendering Methods for Immersive Graph
  Visualization}.
\newblock In {\em Proc. IEEE Pacific Visualization Symposium}, pages 63--67,
  2015.

\bibitem{Kwon16}
O.-H. Kwon, C.~Muelder, K.~Lee, and K.-L. Ma.
\newblock {A Study of Layout, Rendering, and Interaction Methods for Immersive
  Graph Visualization}.
\newblock {\em {IEEE Transactions on visualization and Computer Graphics}},
  22(7):1802--1815, 2016.

\bibitem{Mania01}
K.~Mania and A.~Chalmers.
\newblock {The Effects of Levels of Immersion on Memory and Presence in Virtual
  Environments: A Reality Centered Approach}.
\newblock {\em CyberPsychology \& Behavior}, 4(2):247--264, 2001.

\bibitem{Marriott18}
K.~Marriott, J.~Chen, M.~Hlawatsch, T.~Itoh, M.~A. Nacenta, G.~Reina, and
  W.~Stuerzlinger.
\newblock {\em Immersive Analytics: Time to Reconsider the Value of 3D for
  Information Visualisation}, pages 25--55.
\newblock Springer International Publishing, 2018.

\bibitem{Misue95}
K.~Misue, P.~Eades, W.~Lai, and K.~Sugiyama.
\newblock {Layout Adjustment and the Mental Map}.
\newblock {\em Journal of Visual Languages \& Computing}, 6(2):183--210, 1995.

\bibitem{Newman06}
M.~E.~J. Newman.
\newblock {Finding Community Structure in Networks Using the Eigenvectors of
  Matrices}.
\newblock {\em Physical Review E}, 74:036104, 2006.

\bibitem{North96}
S.~C. North.
\newblock {Incremental Layout in DynaDAG}.
\newblock In {\em Graph Drawing}, volume 1027 of {\em LNCS}, pages 409--418.
  Springer, Heidelberg, 1996.

\bibitem{Purchase07}
H.~C. Purchase, E.~Hoggan, and C.~G\"{o}rg.
\newblock {How Important Is the ``Mental Map''? \textemdash An Empirical
  Investigation of a Dynamic Graph Layout Algorithm}.
\newblock In {\em Graph Drawing}, volume 4372 of {\em LNCS}, pages 184--195.
  Springer, Heidelberg, 2007.

\bibitem{Purchase08}
H.~C. Purchase and A.~Samra.
\newblock Extremes are better: Investigating mental map preservation in dynamic
  graphs.
\newblock In {\em Diagrammatic Representation and Inference}, volume 5223 of
  {\em LNCS}, pages 60--73. Springer, Heidelberg, 2008.

\bibitem{Radu17}
I.~Radu, E.~Southgate, F.~Ortega, and S.~Smith.
\newblock {Summary: IEEE Virtual Reality Workshop on K-12 Embodied Learning
  through Virtual \& Augmented Reality (KELVAR)}.
\newblock In {\em Virtual Reality Workshop on K-12 Embodied Learning through
  Virtual \& Augmented Reality (KELVAR 2017)}, pages 1--2. IEEE, 2017.

\bibitem{Usher18}
W.~Usher, P.~Klacansky, F.~Federer, P.-T. Bremer, A.~Knoll, J.~Yarch,
  A.~Angelucci, and V.~Pascucci.
\newblock {A Virtual Reality Visualization Tool for Neuron Tracing}.
\newblock {\em IEEE Transactions on visualization and Computer Graphics},
  24(1):994--1003, 2018.

\bibitem{Van00}
A.~Van~Dam, A.~S. Forsberg, D.~H. Laidlaw, J.~J. LaViola, and R.~M. Simpson.
\newblock {Immersive VR for Scientific Visualization: A Progress Report}.
\newblock {\em IEEE Computer Graphics and Applications}, 20(6):26--52, 2000.

\bibitem{Watts98}
D.~J. Watts and S.~H. Strogatz.
\newblock {Collective Dynamics of `Small-World' Networks}.
\newblock {\em Nature}, 393:440--442, 1998.

\bibitem{Wickens92}
C.~D. {Wickens}.
\newblock {Virtual Reality and Education}.
\newblock In {\em Proc. IEEE International Conference on Systems, Man, and
  Cybernetics}, pages 842--847, Oct 1992.

\bibitem{Karate}
W.~Zachary.
\newblock {An Information Flow Model for Conflict and Fission in Small Groups}.
\newblock {\em Journal of Anthropological Research}, 33:452--473, 1977.

\bibitem{Zyda05}
M.~Zyda.
\newblock {From Visual Simulation to Virtual Reality to Games}.
\newblock {\em Computer}, 38(9):25--32, 2005.

\end{thebibliography}
